\documentclass[%
 prb,twocolumn,
 superscriptaddress,amsmath,amssymb,
 reprint,%
]{revtex4}

\usepackage{graphicx}% Include figure files
\usepackage{dcolumn}% Align table columns on decimal point
\usepackage{bm}% bold math
\usepackage[utf8]{inputenc}
\usepackage[T1]{fontenc}
\usepackage{mathptmx}
\usepackage{etoolbox}

\usepackage{color}

\newcommand{\rev}[1]{{\color{black}#1}}

\makeatletter
\def\@email#1#2{%
 \endgroup
 \patchcmd{\titleblock@produce}
  {\frontmatter@RRAPformat}
  {\frontmatter@RRAPformat{\produce@RRAP{*#1\href{mailto:#2}{#2}}}\frontmatter@RRAPformat}
  {}{}
}%
\makeatother
\begin{document}

\preprint{AIP/123-QED}

\title[Perspectives on Magnetic/Superconductor Hybrid Systems]{Perspectives on Magnetic/Superconductor Hybrid Systems: 
Long-range Electromagnetic Phenomena Induced by Proximity Effect and Interfacial Spin-orbit Coupling}
% Force line breaks with \\

\author{S. V. Mironov}
\affiliation{Institute for Physics of Microstructures, Russian Academy of Sciences, 603950 Nizhny Novgorod, GSP-105, Russia}

\author{A. S. Mel'nikov}
\affiliation{Institute for Physics of Microstructures, Russian Academy of Sciences, 603950 Nizhny Novgorod, GSP-105, Russia}
\affiliation{Moscow Institute of Physics and Technology (National Research University), Dolgoprudnyi, Moscow region, 141701 Russia}
  
\author{A. I. Buzdin}
\email{alexandre.bouzdine@u-bordeaux.fr}
\affiliation{Institute for Computer Science and Mathematical Modeling,  Sechenov First Moscow State Medical University, Moscow, Russia}
\affiliation{University Bordeaux, LOMA UMR-CNRS 5798, F-33405 Talence Cedex, France}

%\date{\today}

\begin{abstract}
In this Perspective we review recent achievements in physics and applications of hybrid superconductor - ferromagnet structures. In particular, we focus on the manifestations of the electromagnetic phenomena in these systems originating from the response of the induced superconducting correlations modified by the exchange field and  additional  effects coming from the interface Rashba-type spin-orbit coupling. The review includes the long-range electromagnetic proximity effect and related modification of magnetic textures, spontaneous currents, properties of vortex matter and its interaction with magnetic ordering, spin-galvanic, photogalvanic and nonreciprocal transport phenomena in exemplary hybrid systems. We also present our views on the future development of this field including both the \rev{theoretical challenges and promising opportunities for fundamental experiments.} 
\end{abstract}

% fundamental physics and concepts of devices for superconducting spintronics

\maketitle

\section{\label{sec:level1}Introduction}

Superconductivity and ferromagnetism represent two competing types of long-range order: a superconductor expels the magnetic field of a ferromagnet (the Meissner effect), while a ferromagnet, in turn, tends to destroy superconductivity.
There are two primary mechanisms through which magnetic moments and superconducting electrons interact: electromagnetic and exchange mechanisms.
The suppression of superconducting pairing by magnetic induction within a ferromagnet was first addressed by Ginzburg in 1957,\cite{Ginzburg-ZETF56} prior to the development of the BCS theory. By examining the ferromagnetic Meissner state in Type-I superconductors, Ginzburg concluded that coexistence is only possible in the highly specific case where the ferromagnet's magnetic field is compensated by an applied external field. Subsequent studies 
[\onlinecite{Blount-PRL79-OrbitDestr, Ferrel-PRL79-OrbitDestr, Matsumoto-SSC79-OrbitDestr,Krey}] 
further explored the coexistence of these states within the framework of the electromagnetic mechanism.
The second mechanism -- the exchange interaction -- was proposed by Matthias, Suhl, and Corenzwit in the inaugural volume of Physical Review Letters.\cite{Matthias-PRL58,Matt} In a ferromagnet, the exchange field of the magnetic moments acts directly upon the conduction electrons. In a singlet superconductor, the electron spins in a Cooper pair are antiparallel (opposite). However, the exchange field attempts to align these spins in the same direction, thereby destroying the singlet pairing. This phenomenon is known as the paramagnetic effect.
Furthermore, superconducting pairing is suppressed by the exchange scattering of electrons off magnetic moments, as this scattering also disrupts the Cooper pair's singlet state. This was experimentally demonstrated by B. T. Matthias\cite{Matthias-PRL58} and later analyzed in detail by Abrikosov and Gor'kov, who termed the phenomenon magnetic scattering.\cite{AG}
A significant milestone in this field was the 1976 discovery of superconducting single crystals containing a sub-lattice of rare-earth (RE) magnetic atoms, such as RERh$_4$B$_4$, REMo$_6$S$_8$, and REMo$_6$Se$_8$.\cite{Maple} In the majority of these compounds, antiferromagnetism and superconductivity coexist with relatively weak interaction.
However, the behavior differs significantly in ErRh$_4$B$_4$ and HoMo$_6$S$_8$. In these materials, the emergence of long-range ferromagnetic order at a temperature $T_{m}<T_c$ leads to the destruction of superconductivity at a lower temperature, $T_{c2}<T_m$. This reveals a rare re-entrant behavior of superconductivity and a specialized, short-period domain magnetic structure that exists within a very narrow temperature interval where magnetism and superconductivity briefly coexist (for a review, see [\onlinecite{Bulaevskii}]).

A very convenient and attracting possibility to study the 
above interplay between superconducting and magnetic orderings 
appeared with the development of modern nanotechnology which allows to create high quality and controllable 
hybrid structures consisting of superconducting, ferromagnet and normal metal elements.  
For several decades the research activity in this field was focused on two quite different directions: (i) physics of pure electromagnetic 
interaction of the ferromagnetic (F) and superconducting (S) subsystems in the absence of the electron transfer between F and S parts; (ii) proximity phenomena in S/F structures caused by the finite transparency of the interfaces. 
In the latter case the Cooper pairs penetrate into the ferromagnet and
reveal a new physics non-observable in bulk superconductors. 
In particular, the Josephson junction with an F metal as a weak link may transform it into the $\pi$ junction (with the $\pi$ phase difference in the ground state),\cite{Buzdin_Pi, Kupriyanov_pi, Ryazanov, Oboznov, Bobkova_UFN} the
non-collinear magnetization in ferromagnet can generate the triplet
correlations of the superconducting order parameter and the spin-orbit coupling \rev{(SOC)} may result in the $\varphi_0$ Josephson junction\cite{Buzdin_Phi, Reynoso, Zazunov, Mironov_Phi, Kouwenhoven_Phi} with an arbitrary
phase difference $\varphi_0$ in the ground state. Another interesting consequence of the proximity physics in S/F systems is the superconducting spin valve effect which allows to switch the superconducting critical temperature and, thus, the system resistance changing the 
magnetic moment configuration in the F subsystem.\cite{Beasley, Tagirov, Buzdin_FSF, Nowak_1, Leksin_1, Zdravkov_TSVE, Garifullin_rev, Sidorenko_rev,R2_3_1,R2_3_2,R2_3_3}
\rev{Note that the triplet correlations and SOC can manifest themselves in the peculiarities of Andreev reflection at S/F interfaces\cite{Beenakker_1995, R2_2_1, R2_2_2} and enhanced spin-triplet pairing in magnetic junctions.\cite{R2_11_2, R2_11_3} Recent experiments [\onlinecite{R2_12_1}] on the enhanced shot noise in (V/MgO/Fe) junctions reveal possible Josephson effects even with a single superconductor consistent with the proximity-induced Josephson effect proposed by R. Ferrell.\cite{R2_12_2} Moreover, the interfacial SOC leads to the spin valve effect even with a single ferromagnet.\cite{R2_11_1}} Despite obvious fundamental physical interest the work in this direction is stimulated by numerous suggestions of applications  
of these structures in so-called ``superconducting spintronics''. The use of the latter term assumes that the basic working principles
of the related devices are based on various mechanisms of tuning of the supercurrents exploiting the action on the electronic spins. 
All the above phenomena have been observed in experiment in S/F hybrid systems and resulted in the fabrication
of the novel devices quite perspective for the applications in
superconducting spintronics. 
The review of the experimental and theoretical
accomplishments in this domain may be found in Refs.~[\onlinecite{Buzdin-RMP05,Golubov-Kupriyanov-Ilichev-RMP04,Bergeret-RMP05,Eschrig-RPP15,Melnikov,Linder-NatPh15,Amundsen_RMP, R2_1_1, R2_4_1, R2_4_2, R2_4_3}].
The study of physics of electromagnetic interaction between F and S parts of the hybrid structures also provided interesting perspectives for applications even for interfaces nontransparent for electrons. Among them we can mention the control of superconducting transport through   
the creation of tunable superconducting channels along the ferromagnetic domain walls\cite{Buzdin_DWS, Aladyshkin_DWS, Yang_DWS, Gillijns_DWS, Yang_DWS_2, Werner_DWS} and the switching between superconducting vortex states governed by the ferromagnetic elements (see, e.g. [\onlinecite{Aladyshkin_LP}] and references therein).

Considering the obvious success in the developments in these two areas of research of S/F hybrid structures we have to note that these directions cannot be, of course, considered as completely independent. The proximity effect is known to modify strongly the electromagnetic response of the S/F systems and affect, thus, the electromagnetic interaction of S and F elements crucially depends on the exchange field effects. The inverse influence of the electrodynamics can also change the magnetic moment configurations affecting, thus, the proximity phenomena. This mutual influence not only provides the possibility to study exciting new physics but also can open new perspectives in engineering the devices of superconducting spintronics. 

Recently, even deeper connection between the proximity effect and the electrodynamics of S/F hybrids was established with the discovery of the so-called electromagnetic proximity effect (the corresponding term has been introduced in Refs.~[\onlinecite{Mironov-APL-18, Devizorova-PRB-19, Mironov-JETPL-21}]). It was shown that the Cooper pairs penetrating from the superconductor to ferromagnet become involved in the Meissner screening of the magnetization field \textit{inside} the ferromagnet. At the same time, the corresponding currents flowing in the F subsystem become counter-balanced by the dissipationless currents arising inside the superconductor. The latter flow of the superconducting condensate gives rise to the long-range magnetization appearing inside the superconductor. Thus, this effect binds together the exchange and orbital mechanisms of interaction between superconducting and magnetic types of ordering, which becomes the origin for a variety of fascinating phenomena including the spontaneous formation of the current-carrying superconducting states and long-range magnetic field as well as dramatic  modifications of magnetic textures in the ferromagnetic subsystems. 

The above interplay of electrodynamics and proximity phenomena becomes even more spectacular and attractive if we take account of the interfacial \rev{SOC} naturally arising at S/F interfaces. In the absence of the inversion symmetry along a certain direction, the electron momentum becomes coupled with its spin in the ground state. An additional Zeeman-type spin splitting of the electron energy bands due to the exchange interaction between the electron spins with the localized spins of adjacent ferromagnet or external magnetic field lifts the degeneracy of two states with the opposite helicity which result in the appearance of the preferable direction for the electron momentum. As a consequence, S/F systems with strong \rev{SOC} under certain conditions reveal instability towards the formation of spontaneous currents which is accompanied by a variety of nonreciprocal effects. Certainly, the two above mechanisms responsible for the formation of spontaneous current-carrying states may also interfere with each other giving rise to peculiar spin-galvanic phenomena.

It is the goal of the present article to review these new developments in the physics of S/F hybrid systems and discuss the related
perspective functionalities of  superconducting spintronics devices.

We start our discussion of the physics of the mutual influence of the proximity and electromagnetic phenomena in S/F systems from the analysis of electromagnetic proximity effect (Sec.~\ref{sec:EPE}). In sections \ref{sec:level4} and \ref{sec:level3} we analyze possible effects of \rev{SOC} including spontaneous currents, nonreciprocal phenomena and vortex matter properties. In Sec.~\ref{sec:ISGE} we discuss inverse spin-galvanic effect arising in S/F systems with SOC. Some concluding remarks are given in Sec. \ref{sec:level5}.

\section{\label{sec:EPE}  Long-range electromagnetic proximity effect and its experimental signatures}

There are two basic ways how a ferromagnet can transfer the magnetic ordering to the adjacent superconductor. First, its stray magnetic field directly penetrates the superconductor and generates the screening Meissner currents there (orbital mechanism).\cite{Aladyshkin} Second, if there is an electric contact between these two materials the ferromagnet can separate the two electrons forming the Cooper pair by hosting the one with the spin parallel to the exchange field while expelling another one with the opposite spin direction (exchange mechanism). This results in the appearance of small spin-associated magnetization in the surface layer of the superconductor which is often called inverse proximity effect.\cite{Krivoruchko, Bergeret_IPE, Bergeret_IPE_Clean, Lofwander, Faure} From the experimental standpoint, besides very different spatial configurations of the resulting magnetic states the two described phenomena can be distinguished by their typical length-scales. The Meissner currents induced by the orbital effect are rather long-range since they vary at the scale of the London penetration depth $\lambda$ while the spin polarization near S/F interface arises at a distance comparable with the superconducting coherence length $\xi\sim 1-10 ~{\rm nm}$ (roughly, the Cooper pair size).

Recently, it was shown that in addition to the orbital and inverse proximity effects there is another peculiar mechanism of the long-range magnetic field transfer from the ferromagnets to the superconductor in S/F structures -- so-called electromagnetic proximity effect.\cite{Mironov-APL-18, Devizorova-PRB-19, Mironov-JETPL-21, Putilov-PRB-22, Bespalov, Mironov_ISGE, Bobkova_EPE_rev} On the one hand, it arises only if there is an electric contact between S and F subsystems (i.e. in the presence of the proximity effect), but, on the other hand, has purely orbital nature. Specifically, when Cooper pairs penetrate the ferromagnet they become affected by the magnetization-induced magnetic field which results in the generation of the screening Meissner currents flowing \textit{inside} the ferromagnet. Since the total current is zero in the ground state these Meissner currents become counter-balanced by the currents flowing inside the superconductors. The latter ones induce the magnetic field inside the superconductor decaying at distances $\sim \lambda$ from the S/F interface. Thus, although the electromagnetic proximity effect has nothing to do with the stray magnetic field, for typical type-II superconductors with $\lambda\gg\xi$ the induced magnetic field appears to be long-range compared to the spin polarization arising due to the inverse proximity effect. 

%%%%%%%%%%%%%%%%%%%%%%%%%%%%%%%%%%%%%%%%%%%%%
\begin{figure}[hbt!]
\includegraphics[width=0.47\textwidth]{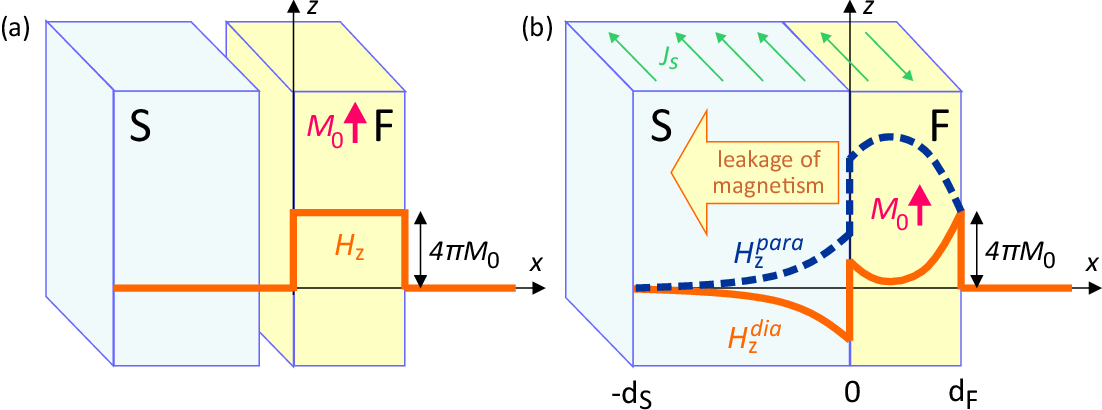}
\caption{Planar
superconductor/ferromagnet system. (a) When the layers are separated from each other the magnetic field exists only inside the F layer. (b) In contrast, when there is electric contact between the layers the magnetization inside the ferromagnet becomes the source of the long-range magnetic field in the superconductor. In both panels the orange solid (blue dashed) curves schematically represent the profile of the magnetic field when the total current inside the F layer is diamagnetic (paramagnetic). \rev{Green arrows schematically show the direction of the superconducting currents flowing inside the S and F layers.} Reproduced from [S. V. Mironov, A. S. Mel'nikov, and A. I. Buzdin, Appl. Phys. Lett. \textbf{113}, 022601 (2018)].}
\label{Fig_EPE_SF}%
\end{figure}
%%%%%%%%%%%%%%%%%%%%%%%%%%%%%%%%%%%%%

To illustrate the main features of the electromagnetic proximity effect let us consider a planar hybrid structure where the superconductor is covered by the ferromagnetic layer with the uniform in-plane magnetization ${\bf M}=M_0{\bf z}_0$  (see Fig.~\ref{Fig_EPE_SF}). In the case when there is no electric contact between S and F layers and the lateral size of this layered system is rather large the chosen direction of magnetization prevents the appearance of the stray magnetic field outside the ferromagnet which ensures that the magnetic field ${\bf B}$ inside the S layer is zero while in the F layer it takes the value ${\bf B}=4\pi{\bf M}$. At the same time, if the S/F interface is transparent for electrons the situation changes dramatically, and the Meissner currents inside the F layer results both in the damping of the magnetic field in the F layer and appearance of the long-range magnetic field ${\bf B}$ inside the superconductor. The specific profile of the magnetic field in both layers is determined by the Maxwell equation for the vector potential ${\bf A}$: 
\begin{equation}
\mathrm{rot}~\mathrm{rot}\mathbf{A}=\frac{4\pi}{c}\left(  \mathbf{j}%
_{s}+\mathbf{j}_{m}\right)  , \label{Eq_Maxwell}%
\end{equation}
where $\mathbf{j}_{s}$ is the Meissner current and $\mathbf{j}_{m}%
=c~\mathrm{rot}\mathbf{M}$ is the magnetization current. The form of dependence ${\bf j}_s\left({\bf A}\right)$ is determined by the ratio between the superconducting zero temperature correlation length $\xi_0$ and the mean free path $l$. In the dirty limit when $l\ll\xi_0$ this dependence has the local form 
\begin{equation}
\label{local}\mathbf{j}_{s}(x)=-\frac{c}{4\pi}\frac{1}{\lambda^{2}%
(x)}\mathbf{A}(x),
\end{equation}
where the London penetration depth $\lambda$ depends on the coordinate $x$ across the bilayer. Note that since Cooper pairs penetrate into the F layer the screening parameter becomes nonzero even inside the ferromagnet and, at the same time, is slightly damped in the layer of the width comparable to $\xi_0$ near the S/F interface in the superconductor. The latter damping can be safely neglected and $\lambda$ can be assumed constant inside the S layer (we will denote this constant as $\lambda_0$) provided the normal state conductivity of the ferromagnet is much smaller than the one in the superconductor. Then assuming that the S and F layer thicknesses satisfy the conditions $d_s\gg\lambda$ and $d_f\ll\lambda$, correspondingly, the solution of Eq.~(\ref{Eq_Maxwell}) in the S
layer takes the form $A_{y}(x)=A_{0}\exp(x/\lambda_{0})$ where $A_{0}$ is a constant
while inside the F layer $A_{y}(x)=A_{0}+4\pi M_{0} x$. As a next step, we integrate Eq.~(\ref{Eq_Maxwell}) over the width of the F layer while using the relation $B_{z}=\partial A_{y}/\partial x$ and accounting that the integral of the magnetization current $\mathbf{j}_{m}$ over the sample thickness is zero. Then the relation between the magnetic field $B_{z}(d_{f})$
outside the sample and the field $B_{z}(0)=A_{0}/\lambda_{0}$ inside the
superconductor close to the S/F interface takes the form
\begin{equation}
\label{integr}B_{z}(d_{f})-B_{z}(0)=A_{0}\int_{0}^{d_{f}}\frac{dx^{\prime}%
}{\lambda^{2}(x^{\prime})}+4\pi M_{0}\int_{0}^{d_{f}}\frac{x^{\prime
}dx^{\prime}}{\lambda^{2}(x^{\prime})}.
\end{equation}
Note that in the limit $d_f\ll\lambda_0$ the first term in the right-hand side of Eq.~(\ref{integr}) can be neglected because its estimate gives the value $\sim(d_{f}/\lambda)B_{z}(0)\ll B_{z}(0)$. In the absence of the external magnetic field, i.e. when $B_{z}(d_{f})=0$, the final expression for the magnetic field inside the superconductor reads
\begin{equation}\label{Eq_result}
{\bf B}=-4\pi M_{0}{\bf Q}\exp(x/\lambda_{0}),
\end{equation}
where the vector kernel ${\bf Q}$ has only the $z$-component which reads 
\begin{equation}\label{Q_def}
Q_z=\int_{0}^{d_{f}}\lambda^{-2}(x^{\prime})x^{\prime}dx^{\prime}. 
\end{equation}
Note that the expression (\ref{Eq_result}) for the induced magnetic field remains the same in the clean limit when $l\gg\xi_0$. In this case the relation between the superconducting current ${\bf j}_s$ and the vector potential ${\bf A}$ becomes non-local, which results only in the renormalization of the expression for the factor $Q$.\cite{Mironov-APL-18} 

To estimate the magnitude of the magnetic field induced inside the superconductor we keep in mind Nb-based systems and take the  magnetization corresponding to typical Fe or Co films $4\pi M_{0}\sim10^{4}~\mathrm{Oe}$. Then assuming $d_{f}$ to be of the order of the superconducting coherence length in the ferromagnet $\xi_{f}$ we find that $Q\sim(\xi_{f}/\lambda)^{2}\sim10^{-2}$ for and, consequently, $B_{z}\sim10^{2}~\mathrm{Oe}$ which is an easily measurable value.

From the experimental standpoint, the most suitable tools for the observation of the electromagnetic proximity effects include spin-polarized neutron reflectometry \cite{Khaydukov-PRB-19, Khaydukov-PRB-14, Khaydukov-JETPL-13} and muon spin spectroscopy.\cite{Flokstra-NatPhys-16, Bernardo-PRX-15, Flokstra-APL-19, Stewart-PRB-19, Flokstra-PRL-18} The main advantage of these techniques is the possibility to reconstruct the spatial profiles of the magnetic field inside the sample and, thus, estimate the typical length-scale of the magnetic field in the S layer. A number of corresponding experiments revealed the presence of magnetic field penetrating into the S layer over anomalously large distances  in V/Fe, Au/Nb/ferromagnet, Cu/Nb/Co and YBaCuO/LaCaMnO hybrids.\cite{Khaydukov-JETPL-13, Flokstra-NatPhys-16, Flokstra-PRL-18, Stahn} These results cannot be explained by the inverse proximity effect since the typical penetration depth values exceed the coherence length $\xi$ up to five times. At the same time, the explanations based on the orbital effect can also be excluded from the consideration since in all structures the magnetization in the F layer has in-plane orientation and the magnetic domains are absent. 

%%%%%%%%%%%%%%%%%%%%%%%%%%%%%%%%%%%%%%%%%%%%%
\begin{figure}[t!]
\includegraphics[width=0.45\textwidth]{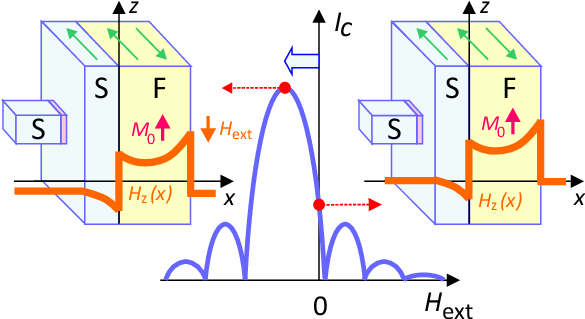}\caption{Shift in the Fraunhofer critical current oscillations for the Josephson junction with one electrode covered by the ferromagnetic layer. Reproduced from [S. V. Mironov, A. S. Mel'nikov, and A. I. Buzdin, Appl. Phys. Lett. \textbf{113}, 022601 (2018)].}%
\label{Fig_CritCurr}%
\end{figure}
%%%%%%%%%%%%%%%%%%%%%%%%%%%%%%%%%%%%%

Also, the electromagnetic proximity effect should reveal itself in transport measurements. Specifically, one may use the S/F bilayer with the S layer thickness $d_s\sim\lambda$ to fabricate the Josephson junction by installing the second small superconducting electrode separated by the insulating barrier from the outer boundary of the large S layer (see Fig.~\ref{Fig_CritCurr}). In this system the Fraunhofer dependence of the critical current on the external magnetic field should become shifted due to the electromagnetic proximity effect. \rev{Recently, an experiment of this type using a Ni/Nb bilayer was reported by N. Satchell et al.\cite{Satchell_SUST} The magnetic field generated by the proximity effect at the outer surface of the $90~{\rm nm}$ thick Nb film was estimated to be below $0.27~{\rm mT}$, which is several times smaller than that observed for the Nb/Co bilayer in [\onlinecite{Flokstra-APL-19}]. 
In other experiments on inverse magnetic hysteresis and Fraunhofer shifts in Nb/Py SIsFS junctions [\onlinecite{Satariano_1}-\onlinecite{Satariano_2}], no conclusive evidence of the inverse electromagnetic effect was found. Instead, the observed Fraunhofer shifts were attributed to the spin polarization of the superconducting layer at the interface with the F layer. Understanding these results remains challenging. One reason may involve a scenario where the superconducting order parameter is strongly damped at the S/F interface, which occurs when the conductivity of the ferromagnet is higher than that of the superconductor above the superconducting critical temperature.} Another possible experiment with the S/F bilayer can be performed using the normal metal tip of the scanning tunneling microscope positioned in the vicinity of the outer boundary of the superconductor. Then the superconducting current arising in the S layer  should change the local density of states due to the Doppler shift of the quasiparticles energy.\cite{Berthod, Karapetrov, MaggioAprile}

Interestingly, in planar S/F/S junctions with in-plane magnetization ${\bf M}$ and coinciding areas of the S and F layers the superconducting currents induced in the superconductors exactly compensate the currents flowing inside the ferromagnetic weak link. As a result, the presence of the magnetization field $4\pi{\bf M}$ inside the F layer and corresponding magnetic flux does not lead to the shift in the Fraunhofer dependence of the critical current vs external magnetic field. \rev{This conclusion shows that in-plane magnetization can not uncontrollably affect the critical current} of Josephson devices used in rapid single flux quantum (RSFQ) logics.

%%%%%%%%%%%%%%%%%%%%%%%%%%%%%%%%%%%%%%%%%%%%
\begin{figure}[t!]
\includegraphics[width=0.48\textwidth]{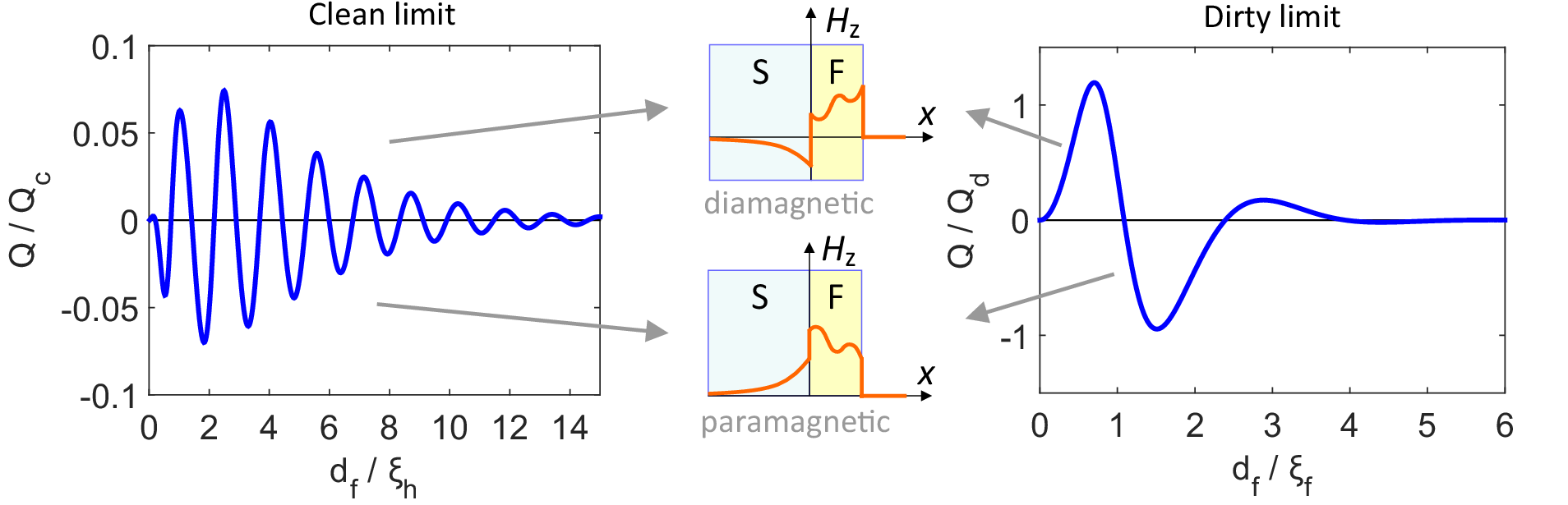}
\caption{The dependence of the
magnetic kernel $Q$ for the S/F bilayer on the F layer thickness $d_f$ in the clean and dirty limits. For the clean limit we take the exchange field $h$ inside the F layer equal to $h=10\pi T$ and choose the temperature $T$ in a way that the superconducting gap $\Delta=2\pi T$. Also we define the values $Q_d=\pi^2\sigma_f\xi_f^2\tanh(\Delta/2T)/(2\hbar c^2)$ and $Q_c=e^2N(0)\xi_h^2(v_F/c)^2$ where $\sigma_f$ is the conductivity of the ferromagnet, $v_F$ is the Fermi velocity, $\xi_h=\hbar v_F/h$ and $N(0)$ is the density of states at the Fermi level per unit spin projection and per unit volume. Reproduced from [S. V. Mironov, A. S. Mel'nikov, and A. I. Buzdin, Appl. Phys. Lett. \textbf{113}, 022601 (2018)].}%
\label{Fig_EPE_Q_df}%
\end{figure}
%%%%%%%%%%%%%%%%%%%%%%%%%%%%%%%%%%%%

Now we turn to the discussion of more subtle phenomena associated with the electromagnetic proximity effect. The first and the most crucial one is the strong oscillatory dependence of the kernel $Q$ in Eq.~(\ref{Eq_result}) on the F layer thickness $d_f$ and the spatial configuration of magnetization. The physics beyond this dependence is closely related to the generation of spin-triplet superconducting correlations inside the ferromagnets.\cite{Bergeret-RMP05} Indeed, for dirty systems the screening parameter $\lambda^{-2}(x)$ entering the expression (\ref{Q_def}) for the kernel $Q$ is determined by the ratio between the singlet ($f_s$) and triplet (${\bf f}_t$) components of the Usadel anomalous Green function $\hat f(x)=\left(f_s+{\bf f}_t \bm{\sigma}\right)i\sigma_y$ (here $\bm {\sigma}$ is the vector of Pauli matrices):\cite{Mironov_FFLO}
\begin{equation}
\label{lambda}
\lambda^{-2}(x)=\frac{16{{\pi }^{2}}T\sigma_N(x) }{\hbar c^2}\sum\limits_{\omega_n >0}{\left[|{{f}_{s}(x)}{{|}^{2}}-|{{{\bf f}}_{t}(x)}{{|}^{2}}\right]}.
\end{equation}
Here $\omega_n=\pi T (2n+1)$ are the fermionic Matsubara frequencies and $\sigma_N(x)$ is the normal state conductivity which may vary from layer to layer (we will denote the corresponding values as $\sigma_s$ and $\sigma_f$ for the S and F layers, respectively). The components $f_s(x)$ and ${\bf f}_t(x)$ are known to reveal damped oscillatory behavior inside the ferromagnets, and the oscillations period is proportional to the superconducting coherence length in the ferromagnet $\xi_f=\sqrt{\hbar D_f/h}$ (here $D_f$ is the diffusion constant in the ferromagnet while $h$ is the exchange field). As a result, if the F layer thickness $d_f$ exceeds $\xi_f$ there are regions where the component $f_t$ dominates over $f_s$ and the screening parameter $\lambda^{-2}$ becomes negative. Consequently, the electromagnetic kernel $Q$ in Eq.~(\ref{Eq_result}) oscillates as a function of $d_{f}$ being either
positive or negative which corresponds to anti-parallel or parallel directions
of $\mathbf{B}$ and $\mathbf{M}_{0}$ for different values of $d_{f}$. The resulting dependencies $Q\left(d_f\right)$ for dirty and clean S/F bilayers are shown in Fig.~\ref{Fig_EPE_Q_df}.   

%%%%%%%%%%%%%%%%%%%%%%%%%%%%%%%%%%%%%%%%%%%%
\begin{figure}[hbt!]\includegraphics[width=0.8\linewidth]{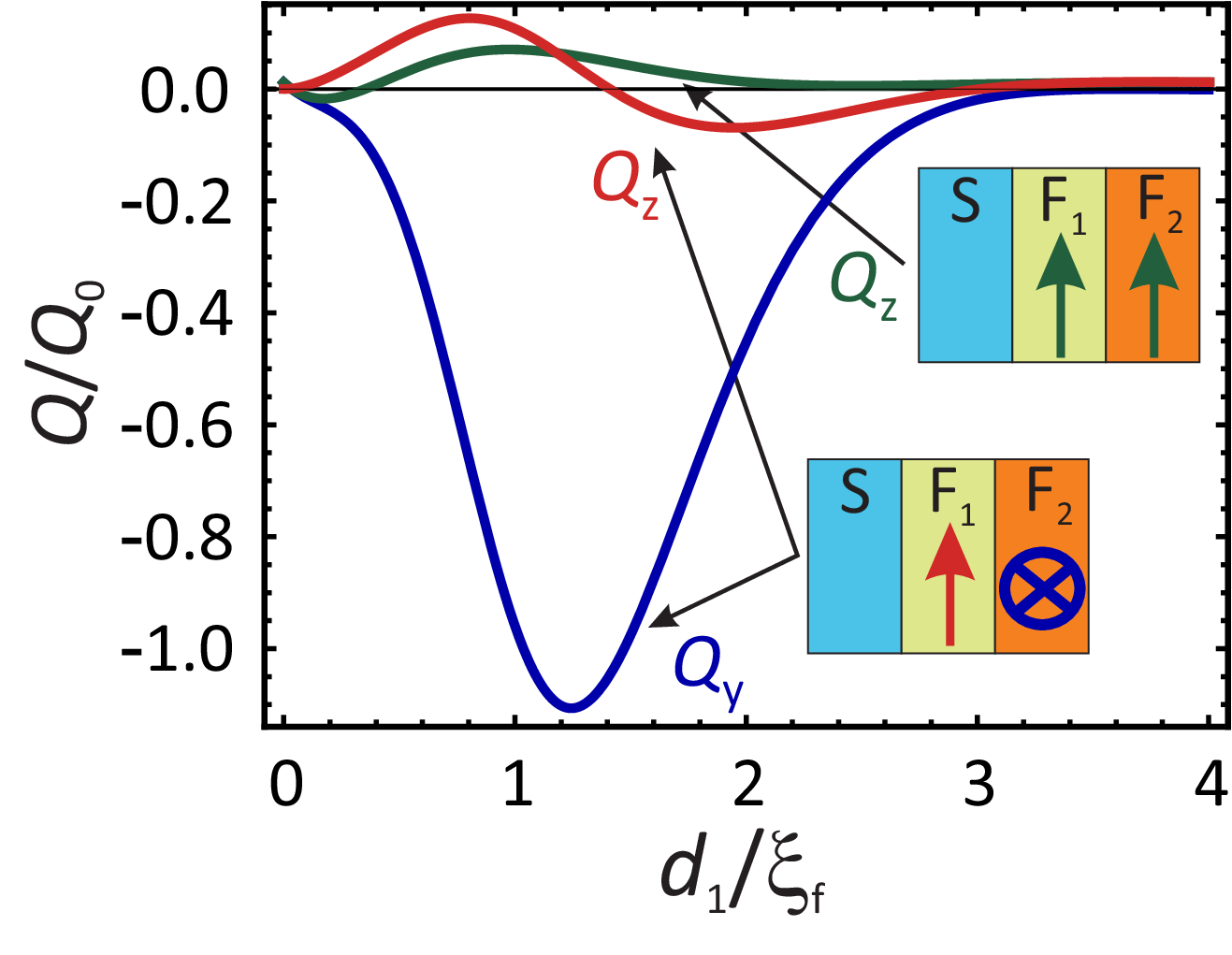}
\caption{The dependencies of magnetic kernels $Q$ for dirty S/F$_1$/F$_2$ structure on the F$_1$ ferromagnet thickness $d_1$. The green curve corresponds to $Q_z$ for the parallel orientation of magnetic moments, while the red and the blue ones are $Q_z$ and $Q_y$ for perpendicular orientation, respectively. Here $Q_0=16\pi^2T\sigma_f \xi_f^2/c^2$ and we take $\Delta=2\pi T$, $h=50\pi T$, $d_2=10\xi_f$. Reproduced from [Zh. Devizorova, S. V. Mironov, A. S. Mel'nikov, and A. Buzdin, Phys. Rev. B \textbf{99}, 104519 (2019)].} 
\label{Fig_EPE_Q_dirty}
\end{figure}
%%%%%%%%%%%%%%%%%%%%%%%%%%%%%%%%%%%%%%%%%%%%%%%%%%

	%%%%%%%%%%%%%%%%%%%%%%%%%%%%%%%%%%%%%%%%%%%%%%%%%%%%%%%%%%%%%%%%%%%%%%%%%%%%%%%%%%%%%%%%%%%
\begin{figure}[hbt!]\includegraphics[width=0.8\linewidth]{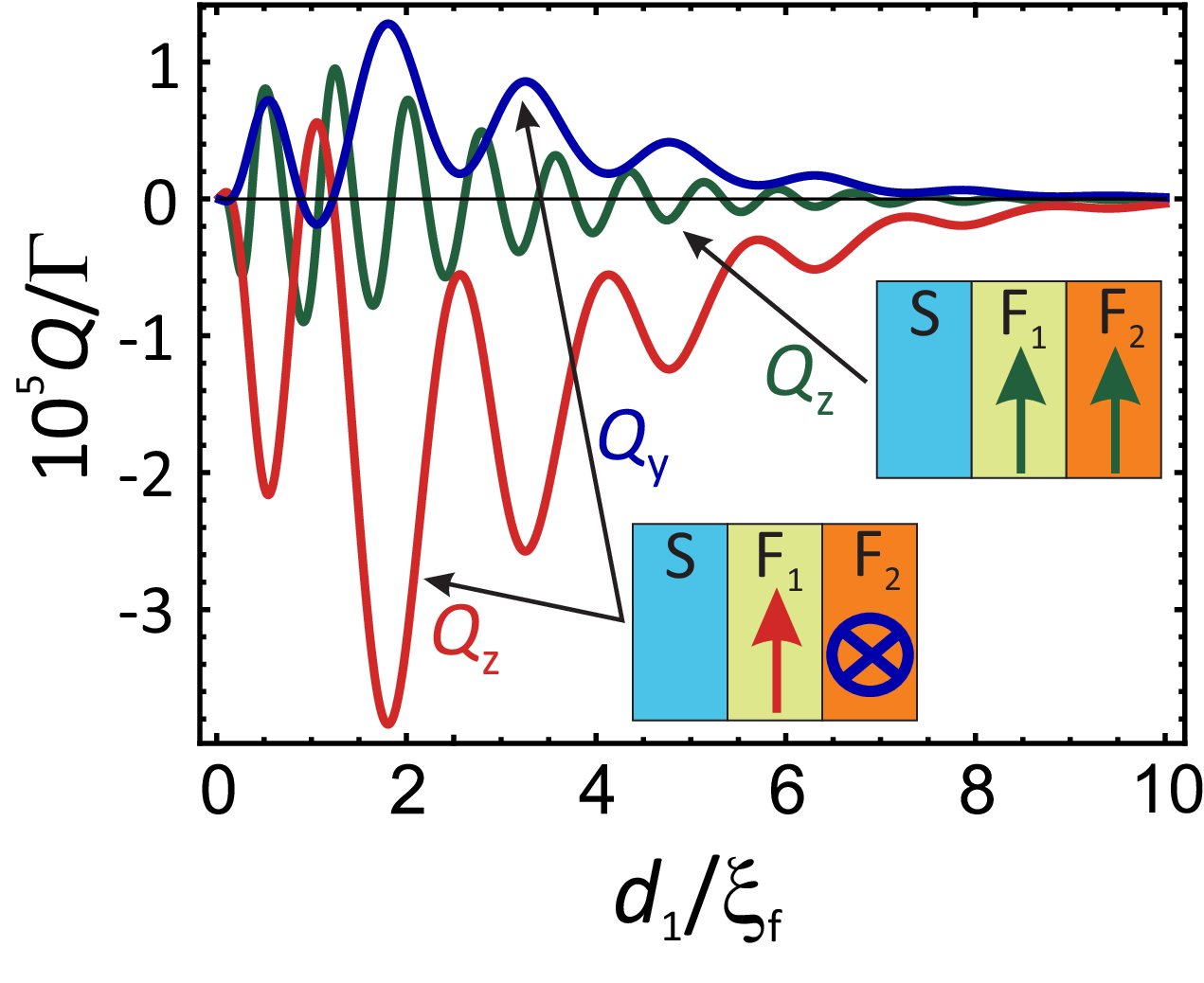}
\caption{The dependencies of magnetic kernels $Q$ for clean S/F$_1$/F$_2$ structure with identical thicknesses of the ferromagnets, i.e. $d_1=d_2$, on $d_1$. Here $\xi_f=\hbar v_F/h$. The green curve corresponds to $Q_z$ for parallel orientation of magnetic moments, while the red and blue ones are $Q_z$ and $Q_y$ for perpendicular orientation, respectively. The parameters are $\Delta=2\pi T$, $h=10\pi T$. Here $\Gamma=8\pi^2 e^2 N(0) v_F^4/(c^2 T^2)$. Reproduced from [Zh. Devizorova, S. V. Mironov, A. S. Mel'nikov, and A. Buzdin, Phys. Rev. B \textbf{99}, 104519 (2019)].}
\label{Fig_EPE_Q_clean} 
\end{figure}
%%%%%%%%%%%%%%%%%%%%%%%%%%%%%%%%%%%%%%%%%%%%%%%%%%

Interestingly, the magnitude of the kernel $Q$ and the induced magnetic field in the superconductor appear to be substantially enhanced provided the magnetization in the F layer forms non-collinear profile.\cite{Mironov-APL-18, Devizorova-PRB-19} The effect is associated to the generation of long-range spin-triplet superconducting correlations which are not destroyed by the exchange field and can penetrate into ferromagnet over the distances comparable to the superconducting coherence length in normal metals $\xi_n=\sqrt{\hbar D_f/T}$. As a result, in the presence of magnetic non-collinearity the region with the well-developed Meissner currents inside the ferromagnetic subsystem becomes substantially broaden which strongly enhance the electromagnetic proximity effect. The basic example is the S/F$_1$/F$_2$ system where the magnetization vectors ${\bf M}_1$ and ${\bf M}_2$ in the two ferromagnetic F$_1$ and F$_2$ layers occupying the regions $0<x<d_1$ and $d_1<x<d_1+d_2$, respectively, form the angle $\theta_M$ with each other so that ${\bf M}_1=M_0 {\bf z}_0$ and ${\bf M}_2=M_0 \left({\bf y}_0\sin \theta_M  + {\bf z}_0\cos\theta_M \right)$. In this case the components of the kernel ${\bf Q}$ in Eq.~(\ref{Eq_result}) take the form \cite{} $Q_z=(Q_1+Q_2 \cos \theta_M+Q_3)$ and $Q_y=Q_2 \sin \theta_M$, where
\begin{equation}\label{Q_def_SFF}
Q_1=\int_0^{d_1} \frac{xdx}{\lambda^2(x)},~~~ Q_2=\int_{d_1}^{d} \frac{(x-d_1)dx}{\lambda^2(x)}, ~~~Q_3=\int_{d_1}^{d} \frac{d_1 dx}{\lambda^2(x)}.
\end{equation}
Note that the screening parameter $\lambda^{-2}(x)$ also depends on $\theta_M$. Further we restrict ourselves to the simplest case when the F$_1$ layer thickness $d_1\sim\xi_f$ while $d_2\gg\xi_n$. When $\theta_M=0$ only the short-range triplet correlations with the zero spin projection appear inside the ferromagnet so that $\lambda^{-2}(x)$ is non-zero in the region of the width $\sim\xi_f$ which gives the estimate $Q_z\sim\left(\xi_f/\lambda_0\right)^2$ discussed above in the context of S/F bilayers. In contrast, for $\theta_M\neq 0$ the appearance of the long-range triplet correlations results in the emergence of another component $Q_y$ of the kernel ${\bf Q}$ and the estimate gives $Q_y\sim\left(\xi_n/\lambda_0\right)^2\gg Q_z$ since for $h\gg T$ (which is typical for experimentally realized systems) one has $\xi_n\gg\xi_f$. As a result, in the experiments where the angle $\theta_M$ can be tuned one should observe a strong increase in the magnetic moment of the S layer for perpendicular orientation of magnetic moments in the two F layers (for $\theta_M=\pi/2$) as compared to the case $\theta_M=0$. Exactly this behavior was observed in experiments with the Au/Nb/ferromagnet structures.\cite{Flokstra-NatPhys-16} Note that this phenomenon should become strongly enhanced in clean S/F$_{1}$/F$_{2}$ structures since in this case the
Cooper pairs penetrate over much larger distances into the ferromagnet and, thus, the region where $\lambda^{-2}(x)$ substantially differs from zero becomes broaden. The typical dependencies of different components of the vector ${\bf Q}$ as a function of the F$_1$ layer thickness for dirty and clean systems are shown in Figs.~\ref{Fig_EPE_Q_dirty} and \ref{Fig_EPE_Q_clean}. 

Note that the enhancement of the spontaneous magnetic field appears also for the in-plane inhomogeneity of magnetization in the ferromagnet.\cite{Bespalov} Specifically, the calculations show that for an infinite slab with two domains and a N\'{e}el wall between them, as well as for a ferromagnetic disk with a magnetization vortex the proximity effect at temperatures below $T_c$ enhances the magnetic field far from the domain wall or the core of the magnetic vortex.

%%%%%%%%%%%%%%%%%%%%%%%%%%%%%%%%%%%%%%%%%%%%%%%%%%
\begin{figure}[hbt!]
\includegraphics[width=0.95\linewidth]{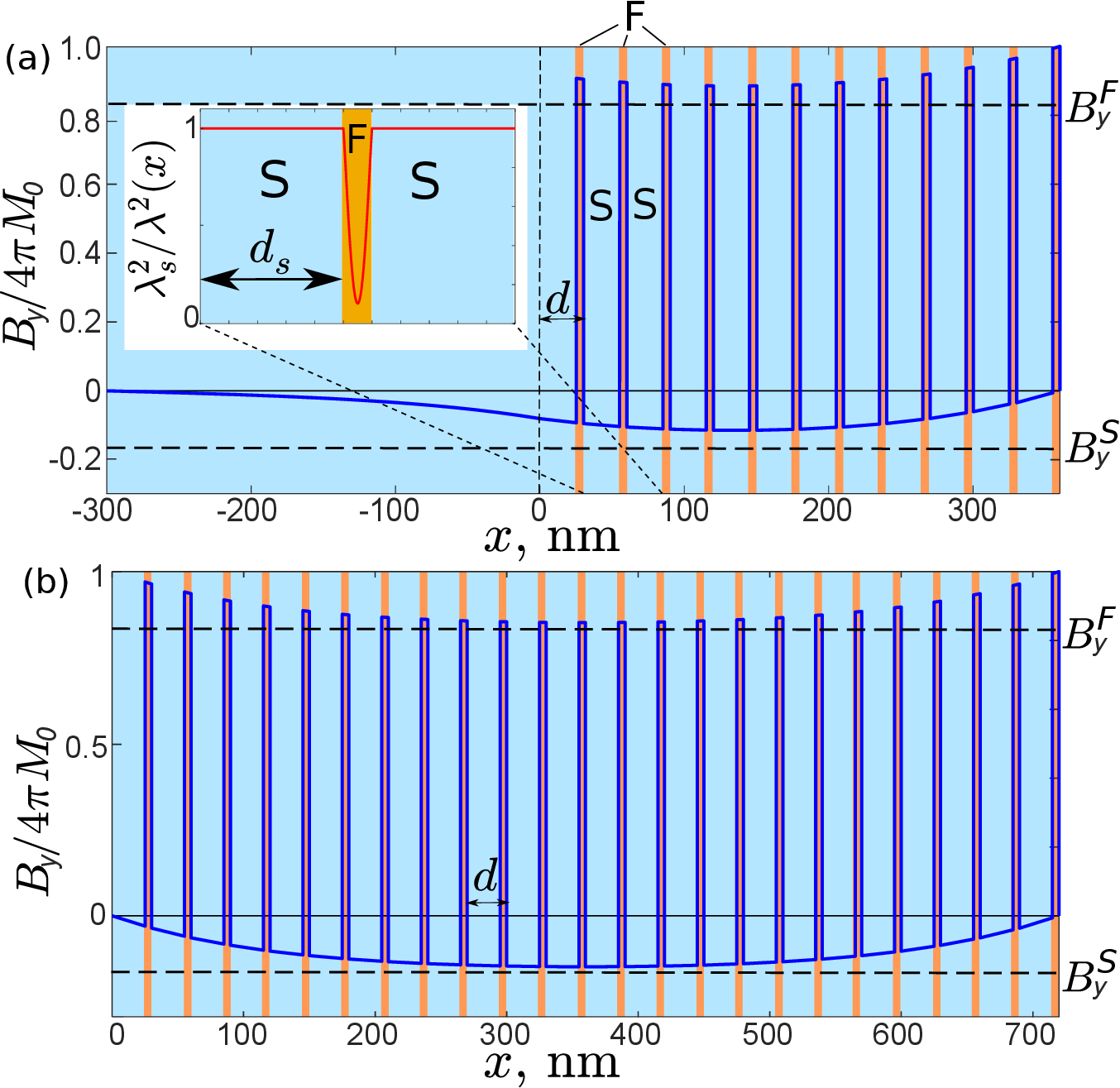}
\caption{The magnetic field profile $B(x)$ in (a) S/F superlattice with $N=12$ bilayers which is in contact with thick ($d_0=300$~nm) superconducting layer ($d_s=25$~nm, $d_f=5$~nm, $\lambda_s=120$~nm). The inset shows the profile $\lambda^{-2}(x)$ inside a single S/F period, $\xi_f=4$~nm. (b) a similar system with $N=24$ bilayers with vacuum or insulator on both outer interfaces. Reproduced from [A. V. Putilov, S. V. Mironov, A. S. Mel'nikov, and A. I. Buzdin, Phys. Rev. B \textbf{105}, 064510 (2022)].}
\label{Fig_EPE_Lattice_B}
\end{figure}
%%%%%%%%%%%%%%%%%%%%%%%%%%%%%%%%%%%%%%%%%%%%%%%%%%

Another possibility to enhance the electromagnetic proximity effect can be realized in S/F superlattices with alternating superconducting and ferromagnetic layers.\cite{Putilov-PRB-22} If the thicknesses $d_s$ and $d_f$ of the S and F layers, respectively, are much smaller than $\lambda$ and the magnetic moments in all F layers are parallel to each other, the S/F superlattice appears to be  similar to the bulk ferromagnetic superconductors with dominating orbital mechanism of magnetic interaction.\cite{Bulaevskii} Deep inside the sample (at distances much larger than $\lambda$ from the sample boundaries) the magnetic field averaged over the period of the superlattice should vanish due to the total compensation of the magnetization field by the Meissner currents. The resulting magnetic field in the S layer reads $B_S=-4\pi M_0d_f/(d_s+d_f)$ while in the F layer it takes the value $B_F=4\pi M_0 d_s/(d_s+d_f)$. Comparing this result with the estimates of the electromagnetic proximity effect in S/F bilayers, one sees that the superlattice may provide the gain $\sim 10^2$ in the magnetic field magnitude which becomes comparable with the magnetization field $4\pi M_0$. Note that if the superlattice is positioned on top of the bulk superconductor the induced magnetic field inside the superconductor would also remain comparable with the value $4\pi M_0$. The profiles of the magnetic field in these two cases are shown in Fig.~\ref{Fig_EPE_Lattice_B}. Interestingly, even for the opposite orientation of the magnetic moments in the neighboring F layers (which is somewhat similar to the anti-ferromagnetic ordering) the magnetic field induced in the adjacent bulk superconductor has the magnitude $\sim \left(d_f/\lambda\right)M_0$ which is larger than the one induced in S/F bilayers.

Experimentally, the profiles of the magnetic field in S/F superlattices were analyzed using neutron scattering measurements.\cite{Khaydukov-PRB-19} The original interpretation of this experiment was based on the model which does not take the electromagnetic proximity effect into account. Specifically, the magnetic field in the F layers was assumed to be defined only by the Meissner screening of the external magnetic field while the cumulative effect coming from the screening of the magnetization field of the F layers was ignored. At the same time, the comparison between the restored magnetic field profiles in Ref.~[\onlinecite{Khaydukov-PRB-19}] and the ones shown in Fig.~\ref{Fig_EPE_Lattice_B}(b) shows that the account of the electromagnetic proximity effect is essential for the accurate interpretation of the experimental data. 

Up to now we have been focused on the magnetic fields induced in the superconducting parts of different S/F system while the magnetic configuration was assumed to be fixed. However, the Meissner currents flowing inside the ferromagnets and subsequently generated magnetic fields interact with magnetic moments in the ferromagnetic subsystem which may give rise to peculiar magnetic reorientation phenomena. Note that the mechanisms responsible for the formation of peculiar domain structures and associated solely with the proximity effect were reviewed in Refs.~[\onlinecite{Buzdin_FFSS}] and [\onlinecite{Buzdin-RMP05}]. Here we complement this review with the recent papers where the electromagnetic proximity effect was shown to substantially modify the magnetic texture in different types of S/F hybrids. The simplest situation is realized in F$_1$/S/F$_2$ systems where the electromagnetic proximity effect favors either parallel or anti-parallel orientation of magnetic moments in two ferromagnets.\cite{Devizorova-PRB-19} \rev{Previously, P. G. de Gennes predicted the anti-ferromagnetic ordering in F$_1$/S/F$_2$ systems with insulating F layers coming from the exchange interaction between the magnetic moments of the F subsystem with the spins of the electrons forming Cooper pairs.\cite{deGennes_FSF} Such peculiar indirect exchange coupling between the F layers was lately observed in experimentally in GdN/Nb/GdN,\cite{R2_7_1, R2_7_2} GdN/V/GdN\cite{R2_7_3} and EuS/Al/
EuS\cite{Li_FSF} trilayers (see also [\onlinecite{Sidorenko_rev}] for review). Here we discuss the alternative mechanism of magnetic ordering in F$_1$/S/F$_2$ systems where the coupling of the F layers arises due to the long-range superconducting currents.} To illustrate the physics beyond this effect we consider the case when the S layer thickness $d_s$ is comparable with $\lambda_0$. Then the mutual orientation of magnetic moments in the ground state corresponding to the minimum of the system free energy per unit area $F_m=\int \left[({\bf B}-4\pi {\bf M})^2/(8\pi) + {\bf A}^2/(8\pi \lambda^2)\right]dx$ can be written in the form $F_m=F_{m0}+\delta F$ where the value $F_{m0}$ does not depend on the angle between magnetic moments and the estimate for the orientation-dependent part $\delta F$ gives 
\begin{equation}
\delta F \sim \frac{\lambda_0  Q_1Q_2}{\sinh(d_0/\lambda_0)} {\bf M}_1 {\bf M}_2,
\end{equation}
where $d_0$ is the total thickness of the structure while $Q_1$ and $Q_2$ are the absolute values of the kernel in Eq.~(\ref{Eq_result}) calculated separately for S/F$_1$ and S/F$_2$ bilayers. In the limit of small thicknesses $d_1$ and $d_2$ of the F$_1$ and F$_2$ layers the kernels $Q_1$ and $Q_2$ are positive which makes anti-parallel configuration of magnetic moments favorable. However, the increase in these thicknesses may change the sign of the kernels making the ground-state angle between magnetic moments an oscillatory function of $d_1$ and $d_2$. \rev{Note that the described mechanism of magnetic interaction through the S layer should be supplemented with the well-known }

\begin{figure*}[hbt!]
\includegraphics[width=0.97\linewidth]{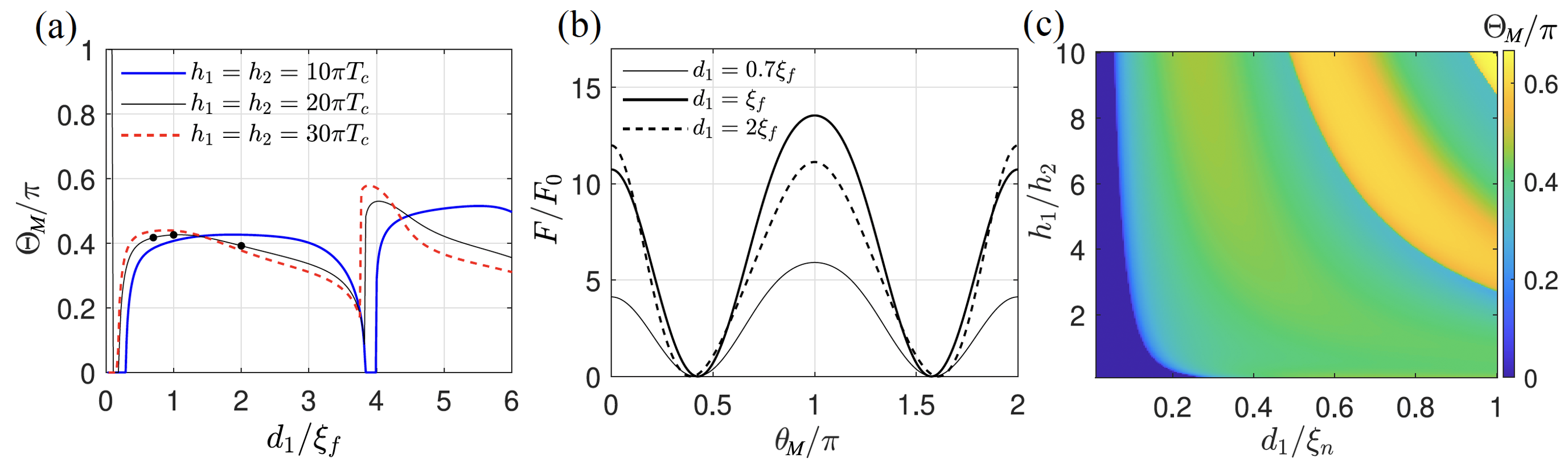}
\caption{(a) Typical dependencies of the ground-state relative angle $\Theta_M$ between the magnetic moments in F$_1$ and F$_2$ layers on the thickness of the $F_1$ layer for $h_1 = h_2 = 10\pi T_c$, $20\pi T_c$, and $30\pi T_c$. Black dots highlight the parameters used in panel (b). (b) Free energy density per unit cross sectional area $F$ vs. the relative angle $\theta_M$ for $h_1 = h_2 = 20\pi T_c$ and $d_1/\xi_f = 0.7$, 1, and 2. (c) Colorplot of $\Theta_M$ as a function of $d_1$ and $h_1/h_2$ for $h_2 = 10\pi T_c$. We choose equal diffusion constants $D$, densities of states at the Fermi level and the inverse coefficients of the molecular field $\alpha_m$ in both ferromagnets defined by the relation $\mu_B M_j=\alpha_m h_j$ ($j=1,2$). Here $F_0 = (4\pi\alpha_m)^2\sigma_N\xi_n D\Delta^2/\mu_B^2c^2$, $\mu_B$ is the Bohr magneton, and we choose $d_2 = \xi_n$ and $\alpha_m k_Fl = 1$, where $k_F$ is the Fermi wave-length and $l$ is the mean free path. Reproduced from [A. A. Kopasov, S. V. Mironov, and A. S. Mel'nikov, Phys. Rev. B, 110, 214501 (2024)].}
\label{Fig_EPE_SFF_magn}
\end{figure*}

Even more interesting situation is realized in the S/F$_1$/F$_2$ systems where the non-collinearity of the magnetic configuration gives rise to the long-range spin-triplet superconducting correlations.\cite{Kopasov_1} In the context of the electromagnetic proximity effect, these triplet correlations (i) substantially broaden the region inside the ferromagnets available for the superconducting correlations and (ii) provide paramagnetic contribution to the screening parameter $\lambda^{-2}(x)$. As a result, the system free energy acquires the terms which depend on the angle $\theta_M$ between magnetization vectors ${\bf M}_1$ and ${\bf M}_2$ in two ferromagnets which effectively corresponds to the emergence of nonlinear interaction between them. This  results in the appearance of several free energy minima as a function of the relative angle $\theta_M$ and their competition with the change in temperature and system parameters. As an outcome, depending on the system parameters the ground state may correspond to parallel, anti-parallel or even non-collinear orientation of magnetic moments. 

The crucial role in this proximity induced reorientation transitions is played by nonlinear contributions in the dependence of the free energy on $\mathbf{M}_1\mathbf{M}_2$ and can't be reduced just to some additional interaction symmetrically equivalent to the Dzyaloshinsky-Moria term (discussed, e.g., in~Ref.~[\onlinecite{MukhamatchinJETPLett2011}]). From the phenomenological standpoint, the free energy contribution describing the interlayer coupling for a ferromagnetic bilayer can be written in the form of the expansion\cite{BrunoPRB1995} $F(\theta_M) = \sum_{n = 0}^{\infty} J_n(\mathbf{M}_1\mathbf{M}_2)^n$, where $J_n$ are certain values which do not depend on magnetic moments and $\mathbf{M}_1\mathbf{M}_2 = M_1M_2\cos\theta_M$. Although the interlayer exchange coupling in non-superconducting ferromagnetic multilayers has been extensively studies starting from seminal works,\cite{GrunbergPRL1986,HeinrichPRL1990,ParkinPRL1990, EdwardsPRL1991,BrunoPRL1991, BrunoPRB1995} not so much attention has been paid to the mechanisms resulting in the nonlinear magnetic interaction beyond the bilinear Heisenberg-type interaction $J_1\mathbf{M}_1\mathbf{M}_2$. The simplest example of such interaction is the biquadratic exchange expressed by the contribution $J_2(\mathbf{M}_1\mathbf{M}_2)^2\propto \cos^2\theta_M$ (see also Ref.~[\onlinecite{DemokritovJPD1998}]). Even this contribution being accounted in the free energy functional results in the competition between the terms $\propto\cos\theta_M$ and $\propto\cos^2\theta_M$ which can qualitatively explain the spontaneous formation of non-collinear magnetic states in a certain range of parameters $J_1$ and $J_2$.

The accurate microscopic calculation of the free energy as a function of the reorientation angle $\theta_M$ shows that surprisingly, in the case when the F$_2$ layer thickness $d_2\gg\xi_f$ the ground-state magnetic configuration is non-collinear in the broad parameter range. The typical dependencies of the ground-state angle $\Theta_M$ on the F$_1$ layer thickness $d_1$ reveal non-trivial oscillatory behavior arising due to the spatial oscillations of the Green's functions inside ferromagnetic layers [see Figs.~\ref{Fig_EPE_SFF_magn}(a)]. Fig.~\ref{Fig_EPE_SFF_magn}(b) shows the evolution of the free energy density per unit cross sectional area profiles for the parameters highlighted by black dots in Fig.~\ref{Fig_EPE_SFF_magn}(a). Finally, Fig.~\ref{Fig_EPE_SFF_magn}(c) represents  the dependencies of the ground-state angle $\Theta_M$ on $d_1$ for different ratios between the exchange field values $h_1$ and $h_2$ in two ferromagnets. Clearly, the non-collinear magnetic states prevail over the collinear ones.  

The magnitude of the proximity induced nonlinear interaction between magnetic moments is zero at $T>T_c$ while below $T_c$ we have $F_m \propto \Delta^2 \propto (1- T/T_c)$. If the magnetic state is collinear at $T_c$, one can expect the emergence of temperature-driven transitions of the magnetic subsystem into a non-collinear state. Such transitions should reveal themselves in peculiarities of magnetization curves, which for a non-collinear magnetic ordering can be sensitive to the orientation of the in-plane external magnetic field. Experimentally, the predicted non-collinearity can be detected by quantum interference device magnetometry,\cite{Lenk} polarized neutron reflectometry,\cite{Sidorenko_reflect} or the muon spin rotation spectroscopy.\cite{Flokstra-NatPhys-16} In addition, changing the orientation of the in-plane magnetic field one can also stimulate the transitions between degenerate non-collinear magnetic states accompanied by hysteresis phenomena. To estimate the effect we take $d_1= \xi_{f_1} = \sqrt{\hbar D_{F_1}/h_1}$, $d_2 \gtrsim \xi_{n_2} = \sqrt{\hbar D_{F_2}/2\pi T}$, and $h_1/h_2  > 1$ and get $\delta F \equiv \max F(\theta_M) - \min F(\theta_M)\sim 10 F_0 \sim 10(4\pi)^2\xi_n N(0)\Delta^2$. For $\xi_n \sim 100$~nm, $\Delta \sim 0.1$~meV and typical density of states at the Fermi level $N(0) \sim 10$~eV$^{-1}$nm$^{-3}$, we find $\delta F \sim 10^{-2}$~eV/nm$^2$. The formation of the non-collinear states driven by the long-range triplet correlations in S/F$_1$/F$_2$ hybrids with thick F$_2$ layer can be also detected by measuring the local density of states (LDOS) at the Fermi level at the outer boundary of the ferromagnetic bilayer. Indeed, if $d_2\gg\xi_f$ only spin triplet correlations should survive at the outer boundary which should result in the increase of LDOS as compared to the similar value in the normal state of the system.\cite{Braude, Buzdin_LDOS, Kontos, Cottet} 

Note that the proximity induced magnetic interaction may also produce magnetic collinearity in S/F bilayers characterized by uniform magnetization at temperatures above $T_c$. In this case below $T_c$ the electromagnetic proximity effect may give rise to the formation of magnetic domains.\cite{Kopasov_2} For the case of a single isolated domain wall the position of the domain wall as well as the resulting magnetic configuration in the F layer appears to be highly sensitive to the thickness $d_f$ of the F layer. For $d_f\ll\xi_f$ the magnetic domains with anti-parallel magnetic moments arise, while in the opposite case the position of the domain wall is shifted towards the S layer and the system features a non-collinear magnetic state. 

Note in conclusion, that since the electromagnetic proximity effect always is an unavoidable consequence of the direct proximity effect its analysis is crucial for the design of the devices of superconducting spintronics. In such systems additional magnetic fields arising from the Meissner currents may substantially affect the operating regimes as well as provide additional functionality. The mutual interaction between long-range electromagnetic effects in S layers and magnetic textures in F layers remains an intriguing open question.

\section{\label{sec:level4} Interfacial Spin-Orbit Coupling: Spontaneous
Currents, Vortices, and Skyrmions}

In this section we discuss another unavoidable consequence of the direct proximity effect in S/F hybrid structures, namely, the influence of the interface Rashba-type \rev{SOC} on the superconducting ordering. 
In general, \rev{SOC} is known to provide the origin of a rich variety of physical phenomena and also serves as an important ingredient of spintronic devices already in nonsuperconducting materials and hybrid structures.\cite{Ivchenko_book, Zutic_RMP} The studies addressing the influence of \rev{SOC} on the superconducting states in noncentrosymmetric crystals have been reviewed in Refs.~[\onlinecite{Mineev_Review, Agterberg_review, R2_8_1}]. In the absence of the inversion symmetry in the direction along a certain unit vector ${\bf n}$ the system energy acquires the term proportional to the scalar invariant $\left({\bf n}\times\sigma\right)\cdot{\bf p}$ where ${\bf p}$ is the electron momentum and $\sigma$ is the vector of the Pauli matrices. As a result, in the ground state the electron spin becomes coupled with the orientation of the momentum ${\bf p}$, and an additional Zeeman spin splitting of the energy spectrum gives rise to the formation of the so-called helical phases with the modulated superconducting order parameter $\Psi({\bf r})\propto
\exp(i\mathbf{qr})$.\cite{Helical0,Helical1,Helical2,Helical3} Recently, the predictions concerning the emergence of helical states appeared to be of particular importance in the context of experimental studies of S/F structures 
with the Rashba \rev{SOC} at the interfaces.
This situation
can be realized, e.g., in a thin superconducting film put on the top of a ferromagnet as the
inversion symmetry is broken and the exchange field is induced in
superconducting layer. The helical superconducting phase reveals a number of interesting properties including  the so called superconducting diode effect (SDE)
recently observed experimentally.\cite{R2_9_1, Diode1,Diode2,Diode3, R2_9_2, R2_9_3}

A very general description of the helical phase may be obtained within the
framework of the modified phenomenological Ginzburg-Landau (GL) theory. As an
example, we may consider a bilayer structure consisting of a thin superconducting
film in contact with a ferromagnet, which serves as a source of the exchange field
acting on the superconducting electrons in the film.
The single-particle Hamiltonian (omitting attraction between electron) is:
\begin{equation}\label{Eq3}
\hat H = \frac{\hat{\mathbf{p}}^2}{2m_e} + \hat{\sigma}_n h_{n} +u_{nm} \hat{\sigma}_n \hat{p}_m \ , 
\end{equation}
where {$\hat{\bf p}$ and $m_e$ are the electron momentum and mass, respectively,} $\hat{\sigma}_{n}$ with $n=x,y,z$ are the Pauli matrices and the summation is assumed over the repeated indices. 
The field $\bf h$ is the
exchange field, the tensor $u_{nm}$  describes the terms linear in the
electron momentum. For example for the Rashba \rev{SOC} at
the interface $u_{nm}=u n_{0s}\epsilon_{snm}$ where tensor $\epsilon_{snm}$ is the Levi-Civita
tensor and the unit vector ${\bf n}_0$ is normal to the interface).

According to [\onlinecite{Mironov_IFE_rev}], the relevant GL gradient terms are
obtained by taking all possible products of pairs of spinsplitting
fields from (\ref{Eq3}) and replacing the momentum $\hat{\bf p}$
with the gauge-invariant gradient. This yields the following
contributions to the odd-gradient part of the GL functional:
\begin{equation}\label{Eq4}
F_{odd}\sim \Psi^*u_{nm} h_{n} \hat{D}_{m} (f_1 +f_3 \hat{\bf D}^2) \Psi \ ,
\end{equation}
where $\hat{\bf D} = -i\nabla - 2e\mathbf{A}/\hbar c$ is the gauge-invarient momentum operator (here $e < 0$).
 The terms with coefficients $f_1$ and $f_3$ describe the joint
effect of the Rashba \rev{SOC} and exchange field.
The full expression for the coefficients  $f_1$ and $f_3$  in (\ref{Eq4}) can
be obtained only on the basis of microscopic theory.
For example in Ref.~[\onlinecite{Plastovets2}] the invariant $F_{odd}$ was derived on
the basis of Gor'kov equations for the Hamiltonian
(\ref{Eq3}) with Rashba-type \rev{SOC}.

For a weak exchange field, $h~< u p_F\ll T_c$ the GL free
energy for a helical state with the wave-vector q is:\cite{Plastovets2}
\begin{equation} \label{GL_q}
F_\text{S}=
\\ 
\Big(a_0+ a_{h} {\bf h}^2 + {\bf q} [{\bf n}_{0}\times{\bf h}_{||}] \big(a_1+a_3 {\bf q}^2\big)+a_2{\bf q}^2  + a_4{\bf q}^4\Big) \Delta^2,
\end{equation}
where ${\bf h}_{||}$ is the in-plane component of the Zeeman field, the coefficients read
\begin{eqnarray} \notag
 a_0= N(0)  \ln\left(\frac{T}{T_c} \right); ~ a_2 =\frac{7\zeta(3)v_F^2N(0)}{32\pi^2T_c^2}; \\  \notag a_4 = \frac{-93\zeta(5)v_F^4 N(0)}{2048\pi^4 T_c^4}; \quad a_1 = u C_u \frac{ \zeta(5) N(0) }{32\pi^4 T_c^2}; \\ \notag  a_3 =-   u C_u  \frac{1905\zeta(7) v_F^2 N(0)}{2048\pi^6 T_c^4}; \quad\quad\quad\quad \\ \label{a_i}
    a_{h} = N(0) \left(\frac{7\zeta(3)}{4\pi^2 T^2_{c0}} - \frac{31\zeta(5)}{32\pi^4}\frac{u^2 p_F^2}{T^4_{c0}} \right)+\mathcal{O}\left(\frac{h^2}{T^2_{c0}}\right),
\end{eqnarray}
the value $C_u$ is defined as  $C_u=u^2p_F^2/T_c^2$  and $N(0)$ is the density of states at the Fermi level. 
Replacing now ${\bf q}$ by  the operator
 $\hat{\mathbf{D}}$ and keeping in mind that $\Delta\propto\Psi$ one can easily restore the form of the gradient terms in Eq.~(\ref{Eq4}).
 We see
that the coefficients $f_1$ and $f_3$ quadratically depend on $u$ in
the considered parameter range. Note that similar results
have been previously obtained in Refs.~[\onlinecite{Helical0}, \onlinecite{edel1}-\onlinecite{Levchenko}].
Interestingly, in the strong SOC regime the dependence of
the coefficients $f_1$ and $f_3$ on $u$ disappears.\cite{edel1, Levchenko}

%%%%%%%%%%%%%%%%%%%%%%%%%%%%%%%%%

The above GL model allows establishing several basic properties of the superconducting systems with SOC. The first one is related to the existence of the spontaneous electric current along the vector $({\bf n}\times{\bf h})$. At a first glance, it seems that the generation of such current is unavoidable because the states with the opposite momenta along the mentioned direction should have different energies. However, the accurate analysis shows that typically no spontaneous current appears. In bulk materials the appearance of the current-carrying states are unfavorable because of the large corresponding kinetic energy of the condensate. This reasoning breaks down for the two-dimensional superconductors subjected to the effect of the in-plane magnetic field, however for these systems the presence of the terms associated with SOC in the GL free energy modifies the expression for the superconducting current in such a way that in the ground state the current is exactly zero. In contrast, for the unconventional d-wave and chiral p-wave superconductors or at the interfaces between the s-wave superconductors and half-metals the appearance of the Andreev edge states leads to the formation of the ground state with broken time-reversal symmetry\cite{Fogelstrom1997, Vorontsov2009, Hakansson2015, Barash2000, Higashitani, Honerkamp, Fauchere, Bobkova, Matsumoto, Stone, Kwon} which may be accompanied by the spontaneous current generation.

%%%%%%%%%%%%%%%%%%%%%%%%%%%%%%%%%%%%%%%%%%%%%%%%
\begin{figure}[hbt!]
\includegraphics[width=0.9\linewidth]{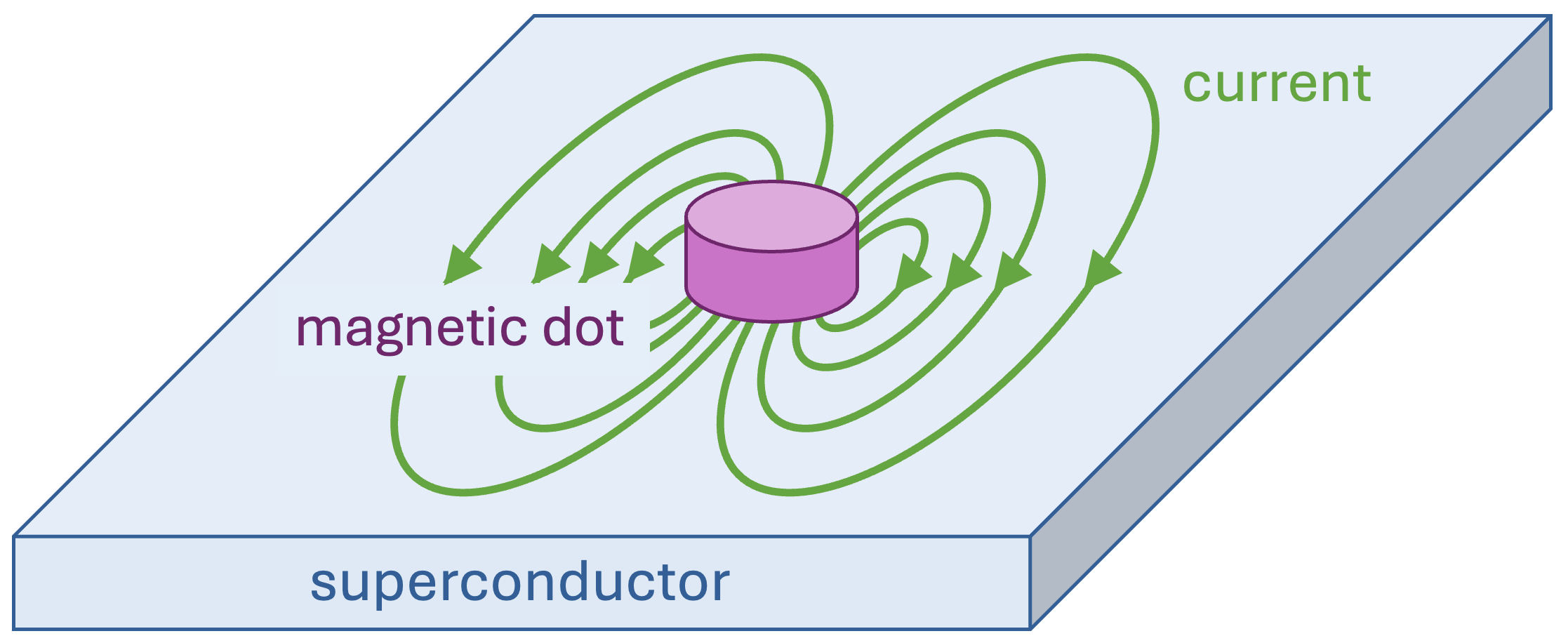}
\caption{The sketch of disipationless currents induced by the small magnetic dot in the underlying superconducting film.}
\label{Fig_magn_dot}
\end{figure}
%%%%%%%%%%%%%%%%%%%%%%%%%%%%%%%%%%%%%%%%%%%%%%%%

Although the helical phase does not generate supercurrents in homogeneous systems, the
situation may be different in the case when the \rev{SOC} is
present only near the small magnetic dots,\cite{Pershoguba} or at the surface
of superconductor.\cite{Mironov-PRL-17,Devizorova-PRB-21} More generally
the non-uniform \rev{SOC} can provide a source of current or phase
difference generation in S/F hybrid which may lead to the new interesting
potential applications.\cite{Baumard,Olde,Robinson2019}
The resulting supercurrents in this case appear either at the boundaries 
between helical and uniform superconducting phases  between the "domains" 
with different momenta of superconducting condensate.
 Between these
regions we get a sort of "domain wall" where the superconducting order parameter must adapt from one phase
to the other. Consequently, a spontaneous superconducting current circulates only
in the vicinity of these domain walls.
This current carrying state was first theoretically considered for the case
of the small magnetic dot at the surface of thin superconducting film in [\onlinecite{Pershoguba}] and later in more details in [\onlinecite{Levitan}]. The sketch of the superconducting currents induced by the magnetic dot is shown in Fig.~\ref{Fig_magn_dot}. The current distribution was calculated neglecting the superconducting screening
which limits the region where current is circulating near the dot by the
Pearl's length.

%%%%%%%%%%%%%%%%%%%%%%%%%%%%%%%%%%%%%%%%%%%%%%%%
\begin{figure}[hbt!]
\includegraphics[width=0.6\linewidth]{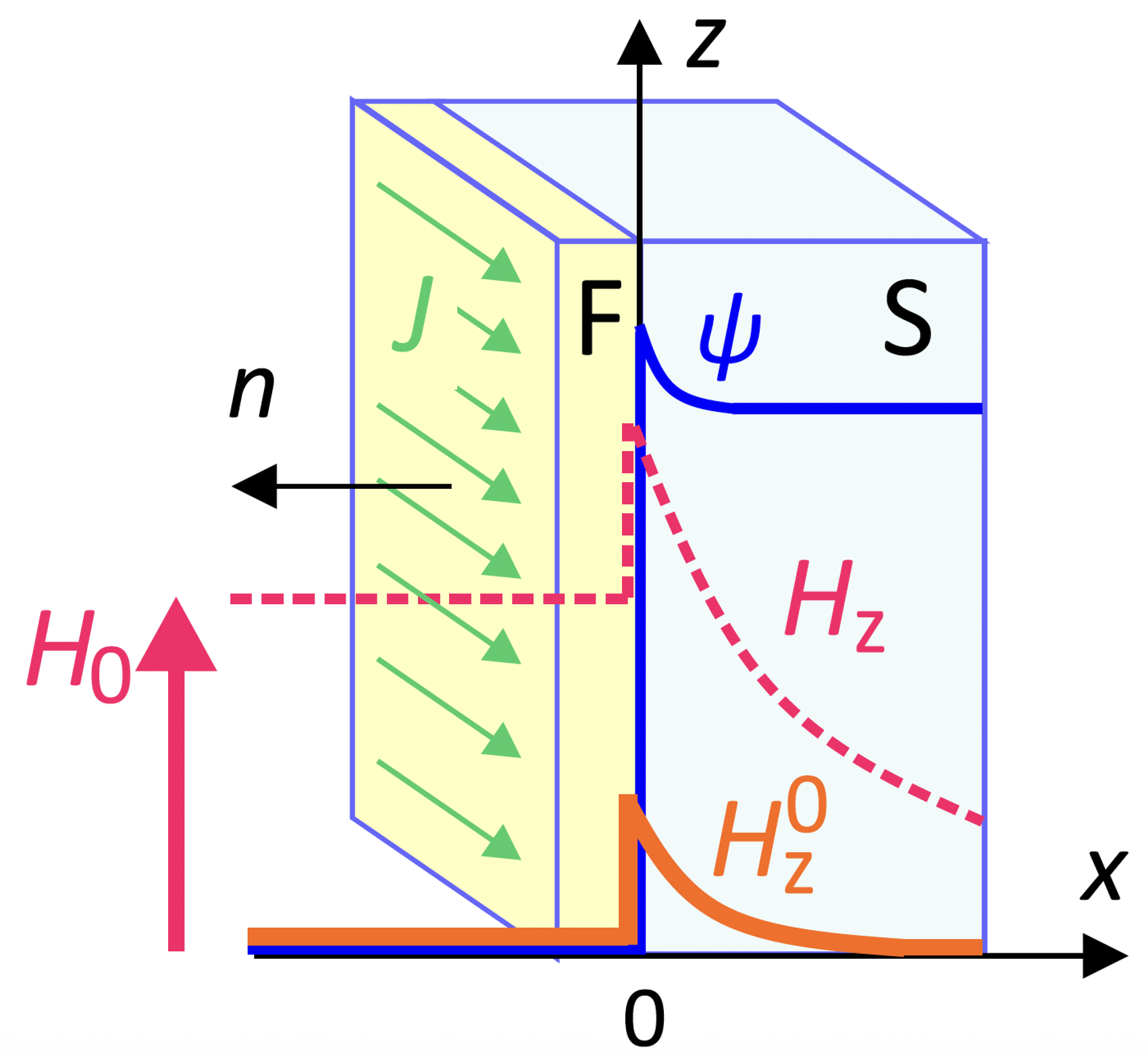}
\caption{Sketch of S/F bilayer with strong SOC which produces spontaneous current ${\bf J}$ (shown with the green arrows). The corresponding profile of the spontaneous magnetic field in the absence of external magnetic field is plotted schematically with the orange color, the distribution of the order parameter $\Psi$ is shown with the blue curve. The magenta curves show the magnetic field profiles in the presence of the external magnetic field ${\bf H}_0$ directed along the $z$-axis. The unit vector {\bf n} is the vector in the direction of the broken inversion symmetry at the S/F interfaces which determines the energy of \rev{SOC}. Reproduced from [S. Mironov, A. Buzdin, Phys. Rev. Lett. \textbf{118}, 077001 (2017)].}
\label{Fig_SF_SOC}
\end{figure}
%%%%%%%%%%%%%%%%%%%%%%%%%%%%%%%%%%%%%%%%%%%%%%%%

Another prominent example of spontaneous current formation occurs in a
two-layer system consisting of a superconductor coated with a ferromagnetic
insulator (FI) layer (the system under consideration is sketched in Fig.~\ref{Fig_SF_SOC}).\cite{Mironov-PRL-17} It was assumed that \rev{SOC} exists only
within a narrow, atomic-thickness layer at the S/F interface. At the same time, the low conductivity of the ferromagnet
ensures that superconducting correlations penetrate the bulk of the
ferromagnet also only at atomic scales. Additionally, it was assumed that the suppression of the
superconducting state due to the proximity effect is negligible so that the modulus of the
order parameter remains independent of the
coordinates throughout the structure.

For simplicity, we considered a configuration where the exchange field $h$ in the
ferromagnet and the external magnetic field $H_0$ are both directed along the
$x$-axis. Under these conditions, the spontaneous current induced by the SOC is
localized near the $z=0$ plane on the atomic scale and flows parallel to the
$y$-axis. Consequently, the SOC-related term in the free energy can be treated
as a surface term
\begin{equation}
F_{so}=2|\Psi|^{2}\varepsilon_1\,l_{so}S\left[  \,\mathbf{n}\times
\mathbf{h}\,\right]  \cdot\left(  \, \nabla\varphi+\frac{\displaystyle2\pi}%
{\displaystyle \Phi_0}\mathbf{A}\,\right)  \bigg\vert_{z=0}\,,
\label{eq2-SOfree_energy}%
\end{equation}
where S is the area of the S/F interface, $\varphi$ is the phase of the superconducting order parameter,$l_{so}$ is the width of the atomically thin region near S/F interface where SOC is sufficient, and $\Phi_{0}=\pi\hbar c/|e|$ is the magnetic flux quantum. In Eq.~(\ref{eq2-SOfree_energy}),
we neglected the spin splitting of electron states associated with the
magnetic field, assuming it to be small compared to the exchange field
($\mu_{B}H_{0}\ll h)$. The surface density of the
spontaneous current $\mathbf{J}_{so}$ flowing along the S/F interface is
determined by the derivative of (\ref{eq2-SOfree_energy}) with respect to the
vector potential,
\begin{equation}
\mathbf{J}_{so}=-\frac{1}{S}\,\frac{c\,\partial F_{so}}{\partial\mathbf{A}%
}=-\frac{c\,\Phi_{so}}{2\pi\lambda^{2}}\,\mathbf{y}_0\,,
\label{eq3-SOsurf_current}%
\end{equation}
where $\Phi_{so}=q_{so}l_{so}\Phi_{0}/4\pi$ \ is the effective magnetic flux, $q_{so}=4m\varepsilon_1
h/\hbar^2$ is the wave number characterizing the spin splitting magnitude in the
presence of SOC and of the exchange field ($q_{so}l_{so}\sim1$),
$\lambda=\left(  mc^{2}/8\pi e^{2}|\Psi|^{2}\right)  ^{1/2}$ is the London
penetration depth and $\mathbf{y}_0$ is the unit vector in the $y$
direction. The above spontaneous surface current leads to the appearance of a
magnetic field jump on the $z=0$ plane. From the Maxwell equations, we find
that the field jump is $\Delta H=2c\Phi_{so}/\lambda^{2}$. The excess
spontaneous field arising in the superconductor must then be screened by the
Meissner currents flowing in the bulk of the S layer.

Estimates show that, for type-II superconductors, the spontaneous field is of
the order of $\Delta H\sim q_{so}l_{so}H_{c1}(T)$ (where\ $H_{c1}(T)=\Phi
_{0}/4\pi\lambda^{2}(T)$\ is the lower critical field of the superconductor)
and for typical values $l_{so}\sim1-10\,\mathrm{nm}$ \ can even exceed the
$H_{c1}(T)$\ field. In this case, the spontaneous current can make it
energetically favorable for Abrikosov vortices to enter the sample.
We first discuss the limit of a thick superconductor, with the S layer
thickness $L\gg\mathrm{max}\left\{  \lambda,\xi\right\}$.\cite{Mironov-PRL-17} The Meissner currents screening the spontaneous current
$\mathbf{J}_{so}$ are then localized on the $\lambda$ scale near the S/F
interface, and the magnetic field profile inside the superconductor has the
form $H_{x}(z)=(H_{0}+\Delta H)\exp(-z/\lambda)$.

In type-II superconductors, the spontaneous current can also affect significantly the
temperature dependence of the critical magnetic field $H_{c3}$, that
corresponds to the appearance of localized superconductivity at fields
exceeding the bulk critical field. This effect is a consequence of the
interaction of the spontaneous current with the Meissner current that screens
the external field $H_{0}$. Calculations presented in Ref.~[\onlinecite{Mironov-PRL-17}] show that,
at temperatures close to $T_{c}$, the dependence $H_{c3}(T)$ \ has the form
\begin{equation}
H_{c3}(T)\approx H_{c3}^{0}(T)\left[  1\pm\zeta\Phi_{so}/\Phi_{0}\right]  ,
\label{SOC_Hc3}%
\end{equation}
where $H_{c3}^{0}(T)=1.69(2ma/e)(T_{c}-T)$ is the temperature dependence
of the critical field for surface superconductivity in an isolated S layer,\cite{deGennes-book} $a \propto(1-T/T_c)$ is the standard GL coefficient, and $\zeta\approx24.9$ is a constant. The upper (lower) sign in the above expression corresponds to the coinciding (opposite) directions of 
 the exchange field in the F layer and the external magnetic field. Hence, a change in the direction of the external
magnetic field should lead to a change in the slope of the $H_{c3}%
(T)$\ dependence, which can be used as a tool for the experimental
detection of spontaneous currents.

%%%%%%%%%%%%%%%%%%%%%%%
%%%%%%%%%%%%%%%%%%%%%%%
\begin{figure}[hbt!]
\begin{center}
\includegraphics[width=0.5\linewidth]{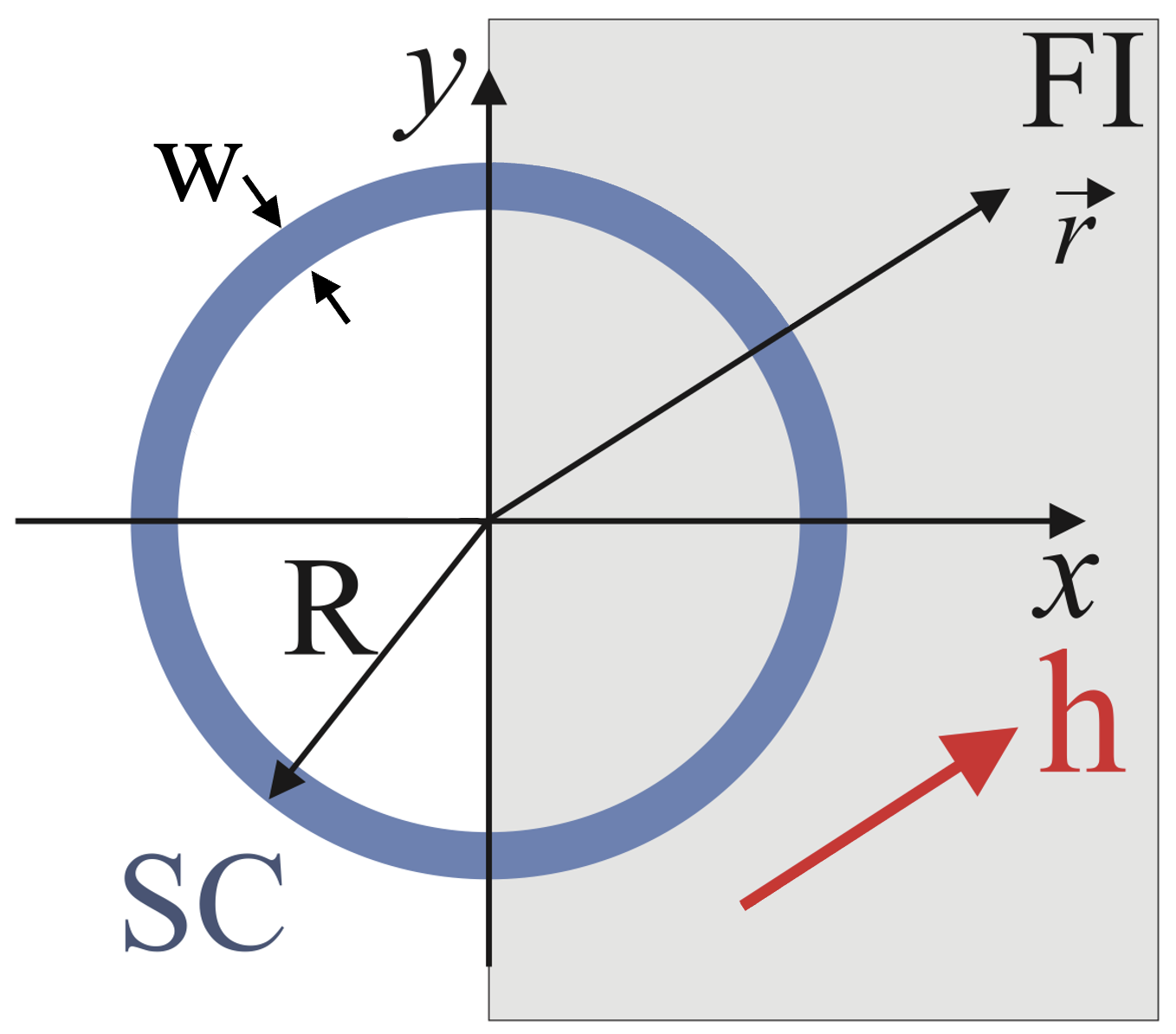}
\end{center}
\caption{Sketch of the thin-film superconducting loop proximitized with a ferromagnetic insulator (FI) occupying the half plane $x>0$. Reproduced from [J. W. A. Robinson, A. V. Samokhvalov, A. I.
Buzdin, Phys. Rev. B \textbf{99}, 180501(R) (2019)].}\label{Fig_ring_SOC}
\end{figure}
%%%%%%%%%%%%%%%%%%%%%%%
%%%%%%%%%%%%%%%%%%%%%%%

Another way to detect the spontaneous supercurrent induced by the combined
action of the Rashba SOC and the exchange field at an S/F interface can
be implemented in systems with a multiply connected geometry,\cite{Robinson2019} for example, in a superconducting circuit partially coated with a ferromagnetic insulator (see Fig.~\ref{Fig_ring_SOC}). Assuming that the circuit is thin and narrow, $d,\,w\,\ll\xi_{0}\sim\hbar v_{F}/\Delta_{0}$, and the
temperature $T$ is considerably lower than $T_c$, we can regard the amplitude of
the wave function of the pairs as being homogeneous along the circuit
($|\Psi|=\mathrm{const}$). The presence of the SOC and the exchange
field (averaged over the small thickness $d$ of the superconducting wire)
gives rise to the phase modulation $\Psi=|\Psi|e^{iq_{1}s}$ with the wave
vector $q_{1}$ in a length-$L_{h}$ part of the circuit oriented along the
vector $[\mathbf{n}\times\mathbf{h}]$ which is in this case directed along the
$y$-axis ($s$ is a coordinate along the circuit and the Rashba vector
$\mathbf{n}$ is orthogonal to the circuit plane). The corresponding
supercurrent density
\begin{equation}
j=\frac{e\hbar}{m}|\Psi|^{2}\left(  \frac{\partial\varphi}{\partial s
}-q_{so}\right)
\end{equation}
in this region is $j_{1}=(e\hbar/m)\,|\Psi|^{2}(q_{1}-q_{so})$. In the
remaining length- $L_{0}$ part of the circuit, the SOC and exchange
field are both absent ($q_{so}=0$), and the supercurrent density is determined
by the usual expression $j_{2}=(e\hbar/m)\,|\Psi|^{2}q_{2}$ where the wave
number $q_{2}$ is to be found from the current continuity condition
$q_{2}=q_{1}-q_{so}$ and the single-valuedness of the wave function phase
$q_{1}L_{h}+q_{2}L_{0}=2\pi n$. From these conditions, it is easy to determine
that the wave number $q_{2}=(2\pi n-q_{so}L_{h})/(L_{h}+L_{0})$\ is nonzero in
general and this means the possibility of spontaneous generation of
supercurrent in the circuit. Thus, a superconductor with an exchange field and
SOC can under certain conditions play the role of a phase battery that
produces the phase difference $\triangle\varphi=q_{1}L_{h}$ in part of the
circuit.\cite{Robinson2019} A tunable phase battery based on a combination
of SOC and exchange interaction in an InAs nanowire was recently
implemented in Ref.~[\onlinecite{Strambini}].

We note that the above spontaneous generation of supercurrent is not specific
to S/F hybrids with the Rashba SOC, but should be relevant for a wide
class of interfaces between superconductors and materials with spin
polarization and broken spatial inversion symmetry, such as topological
insulators and materials with full spin polarization.

As it was mentioned the current may appear at the interface between helical
and normal superconductors. Interestingly if this current is large enough it
can generate the Abrikosov vortices.
The most suitable geometry for this seems to be a thin superconducting film
with a small thickness $d\sim\xi_{0}$ partially covered by a thin-film
metallic ferromagnetic strip (of width $L$) with in-plane exchange field ${\bf h}$ and strong interfacial Rashba SOC.\cite{Olde} Introducing the effective
magnetic flux characterizing the SOC $\Phi_{so}=q_{so}l_{so}\Phi
_{0}/4\pi$ and taking into account its spatial dependence $\Phi_{so}
(\mathbf{r})$ we can write the superconducting free energy of our
effectively 2D system with a vortex in the London approximation as
\begin{equation}
F=\frac{\lambda_{eff}^{-1}}{8\pi}\int\left[  \left(  \Phi-\mathbf{A}\right)
^{2}+\frac{4\Phi_{so}(\mathbf{r})}{d}{\bf y}_0\cdot\left(\Phi-\mathbf{A}\right)  \right]
d^{2}\mathbf{r+}\int\frac{B^{2}}{8\pi}d^{3}\mathbf{r}\label{SOFunct}%
\end{equation}
where ${\bf y}_0$ is the unit vector along the vector product ${\bf n}\times{\bf h}$, the presence of vortex at the point $\mathbf{r}_{0}$ , following
Ref.~[\onlinecite{deGennes-book}], is described by the vector quantity $\Phi$ proportional to the gradient of superconducting phase
\begin{equation}
\Phi=\Phi_{0}\frac{\mathbf{n}\times\left(  \mathbf{r}-\mathbf{r}%
_{0}\right)  }{\left(  \mathbf{r}-\mathbf{r}_{0}\right)  ^{2}}.%
\end{equation}
Here $\lambda_{eff}=\lambda^{2}/d$ is a Pearl's length describing the
screening in a thin film limit $\lambda>>d$.
Evaluating the  current
distribution \cite{Olde} we find the condition of spontaneous vortices generation 
\begin{equation}
\Phi_{so}>\frac{\Phi_{0}}{16}\frac{d}{\lambda_{eff}}\frac{\ln(\lambda
_{eff}/\xi)}{\ln(L/\lambda_{eff})}.
\end{equation}
The estimates presented in Ref.~[\onlinecite{Olde}] show that for the films with a small
$d/\lambda_{eff}$ ratio the spontaneous generation of vortices should be
easily realized.  This phenomenon of spontaneous vortex generation for the S/F systems with
different forms of magnetic regions appears to be a quite general phenomenon and
has been studied for different  geometries in [\onlinecite{Samokhvalov}] and [\onlinecite{Malshukov}].
It is obvious that an analogous mechanism of generation of
superconducting currents and Abrikosov vortices via SOC in S/F bilayers can be realized for different types of
 localized magnetic textures  such as
magnetic domain walls or N\'{e}el skyrmions. Using the functional
(\ref{SOFunct}), where the value $\Phi_{so}(\mathbf{r})$ corresponds to the exchange field
generated by the skyrmion we may find the superconducting current distribution
and the condition of spontaneous vortices generation.\cite{Baumard} Note in this context that recently
there appeared interesting experimental evidences of vortex generation
by the skyrmions in S/F hybrids with thin insulating barrier at the S/F
interface.\cite{Petrovic,Xie} The mechanism of vortex-skyrmion
interaction in this case is purely electromagnetic one and the reader can find
a review of different theoretical models describing this situation in Ref.~[\onlinecite{Apostoloff}].

Another intriguing question associated with the physics of S/F hybrids with SOC is related to the so-called diode effect. Typically, this term means that the critical depairing current of the superconducting system depends on the direction of the current. One may recognize at least two basic mechanisms responsible for the appearance of the diode effect due to SOC in the GL theory. The first one is associated with the terms linear over the momentum ${\bf q}$ in the GL free energy functional (\ref{GL_q}) while another one is related to the nonlinear terms. The effect of the linear terms on the anisotropy of the critical current arises, e.g., in S/F bilayers with thin superconductors of the width $L\sim\xi$ and strong
SOC considered in Ref.~[\onlinecite{Devizorova-PRB-21}]. It was was shown that in such planar system  the deparing current depends on the current direction in the plane of the sample.
Such an anisotropy of the critical current does not seem possible at first
glance, because the total superconducting current due to the SOC and
the Meissner effect are equal to zero. But the key point is the spatial
inhomogeneity of the distribution of the Meissner current that screens the
$\mathbf{J}_{so}$ current along the thickness of the S/F structure. As a
result, for a fixed orientation of the exchange field, the local current
density (and hence local suppression of the superconducting order parameter)
at some point in the superconductor becomes dependent on the angle 
between the external transport current and the spontaneous current
$\mathbf{J}_{so}$ flowing along the S/F interface. The maximum current that can
flow along the S/F structure without destroying superconductivity also acquires
dependence on this angle. The described diode effect allows detecting
spontaneous currents in transport measurements. Note, that recently it was experimentally demonstrated that the critical temperature and the critical current of the thin elongated superconducting stripe made of
aluminum and placed on top of dielectric ferromagnet yttrium iron garnet (YIG)
with in-plane magnetization depend on the mutual orientation between the
current flow and ferromagnet magnetization.\cite{Uspenskaya2021} The analysis
provided in Ref.~[\onlinecite{Putilov2024}] shows that the explanation of this effect may be
related to the nonreciprocal electron transport in thin
superconductor/ferromagnet bilayers with strong \rev{SOC}.

\begin{figure}[t!]
    \centering
    \includegraphics[width=0.9\linewidth]{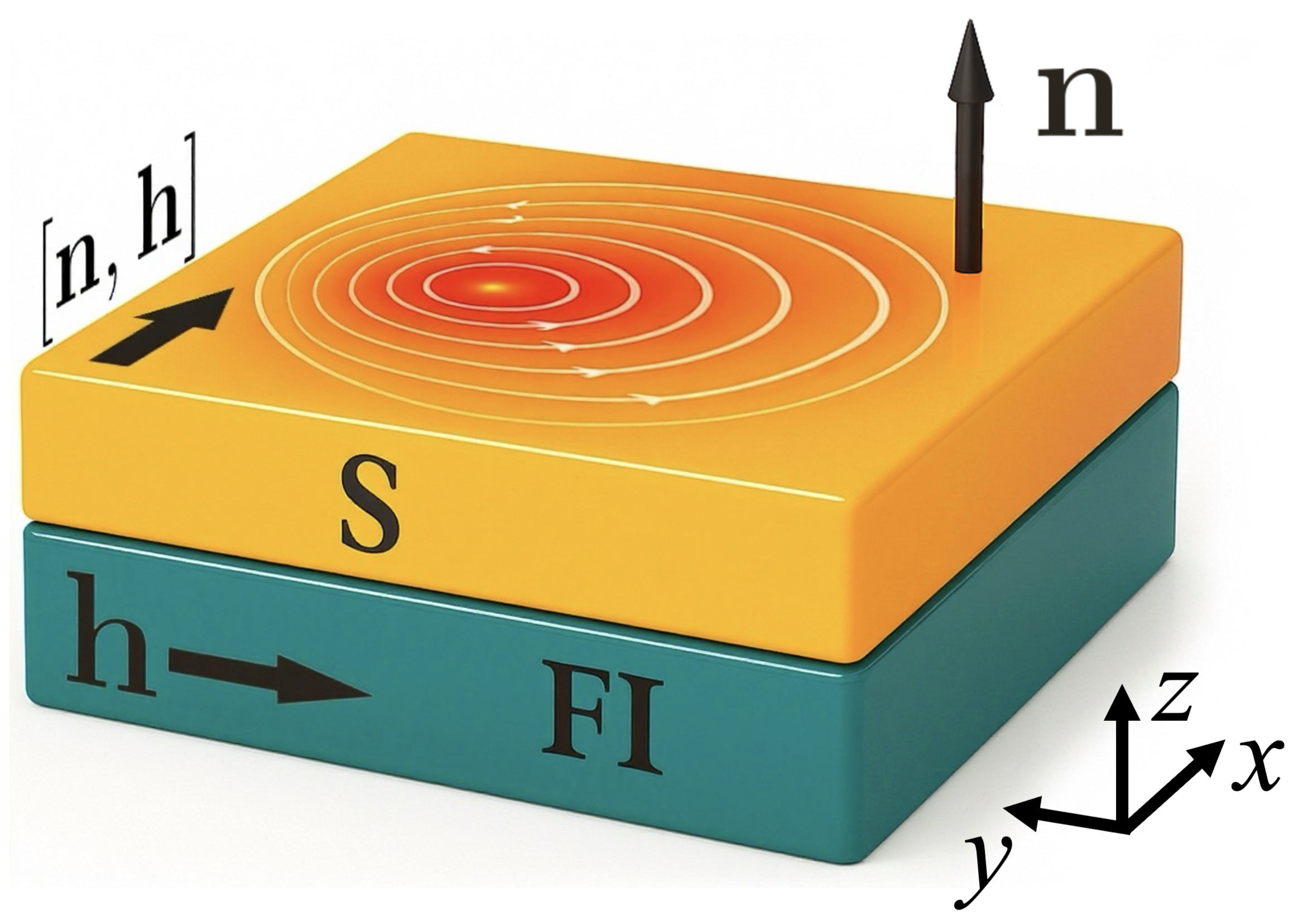}
    \caption{A thin superconducting layer (orange, S) rests on a ferromagnetic insulator (teal, FI).  The ferromagnet’s uniform in‐plane exchange field $\mathbf{h}$ and the interface normal $\mathbf{n}$ (along $z$) produce the intrinsic direction $[\mathbf{n}\times\mathbf{h}]$ in the $xy$-plane.  In the superconductor we depict the circular supercurrent of a single vortex, whose profile is deformed by the cubic \rev{SOC} term aligned with $x$.  The axes $x$, $y$, and $z$ indicate the laboratory frame; arrows show the directions of $\mathbf{h}$, $\mathbf{n}$, and the vortex circulation. Reproduced from [A. V. Putilov, D. V. Zakharov, A. Kudlis, A. S. Mel'nikov, A. I. Buzdin, Phys. Rev. B \textbf{112}, 134507 (2025)].}
    \label{fig:skt}
\end{figure}

The second mechanism underlying the diode effect is associated with the terms in the GL functional which are nonlinear with respect the momentum operator. This situation is often referred as intrinsic diode effect. One can naturally expect that
the intrinsic diode effect should cause drastic modifications and provoke intriguing new phenomena in the properties of vortex matter, including structure of individual vortices and vortex lattices as well as the vortex dynamics.\cite{Putilov2025}
Recent experiments on measurements of the vortex-induced kinetic inductance in meander microstrips with Rashba-type interaction
indicate such modification of vortex matter properties.\cite{Fuchs-PRX-22} These studies motivated us to address the problem 
of non-reciprocal phenomena in the vortex state and
suggest a theoretical model that describes the structure of individual vortices and vortex interactions in the low magnetic field regime, i.e., for low vortex concentrations.\cite{Putilov2025}  
We considered some generic exemplary system, namely, a thin superconducting film deposited on an \emph{insulating} ferromagnet. In Fig.~\ref{fig:skt}, we schematically present this structure. The interface normal $\mathbf n$ breaks the inversion symmetry, while an in-plane exchange field $\mathbf h$ breaks the time-reversal symmetry. Because the ferromagnet is insulating, the inverse proximity effect and related suppression of superconducting order parameter are negligible. 
%The thickness $d$ of the superconducting film must be relatively small to make surface effects more pronounced, and the vectors $\mathbf n$ and $\mathbf h$ together single out the in-plane direction $[\mathbf n\times\mathbf h]$ that governs all non-reciprocal responses discussed below. 
Hereafter we neglect the Zeeman field effect, which can also disturb the vortex magnetic field in structures with broken inversion symmetry.\cite{Agterberg-PRB-07,Lu-PRB-08,Lu-JLTP-09,Yip-JLTP-2005}
%As discussed above the spin-orbit coupling at the superconductor/ferromagnet interface enters the GL theory through the odd spatial-derivative terms. Averaging these contributions across a film only a coherence-length thick produces an effective two-dimensional free-energy functional that augments the usual GL terms as well as the linear and cubic gradient contributions responsible for the intrinsic diode effect. The linear spin-orbit term forces the equilibrium condensate to form a helical pattern whose wave vector points along the cross-product of the interface normal and the in-plane exchange field; the cubic gradient term gives the asymmetry in the nonlinear relation between the current density and superfluid velocity giving, thus, the nonreciprocal response. 

%%%%%%%%%%%%%%%%%%%%%%%%%%%%%%%%%%%%%%%%%%%%%%%%
%%%%%%%%%%%%%%%%%%%%%%%%%%%%%%%%%%%%%%%%%%%%%%%%

As discussed above the exchange field and the interfacial \rev{SOC} can be included into the GL free-energy density through the odd-gradient terms (see also Refs.~[\onlinecite{Mineev-JETP-94}, \onlinecite{edel1}, \onlinecite{Helical2}, \onlinecite{Mironov-PRL-17}, \onlinecite{Daido-PRL-22}]).  
Considering the SOC confined to an atomic-scale interface layer \cite{Mironov-PRL-17} taking rather small film thickness $d$ (less than coherence length $\xi$) we neglect the variations of the order parameter along $z$ and average the free energy density over $d$:
\begin{equation}
\begin{array}{c}{\displaystyle
F= -a|\Psi|^{2}
          +\frac{b}{2}|\Psi|^{4}
          +\xi^{2} a|{\bf D}\Psi|^{2}\label{Eq_GLfunc} } \\{}\\{\displaystyle 
          +\left(\varepsilon_{1}\Psi^{*}[\mathbf n\!\times\!\mathbf h]\mathbf D\Psi
          +\varepsilon_{3}({\bf D}\Psi)^{*}[\mathbf n\!\times\!\mathbf h]\mathbf D^{2}\Psi
          +\text{c.c.}\right),}
\end{array}
\end{equation}
where $a \propto(1-T/T_c)$, $b>0$ are the standard GL coefficients, $T_c$ is the superconducting critical temperature and operator $\mathbf D=-i\nabla-2\pi\mathbf A/\Phi_0$ where $\mathbf A$ is a vector potential.  
%The phenomenological coefficients $\varepsilon_{1}$ and $\varepsilon_{3}$ quantify the strength of the linear and cubic gradient terms, respectively. The higher order odd and even gradient terms are omitted as we consider large scale spatial variations of the order parameter in the plane of the 
%superconducting layer.
The order parameter takes the form $\Psi=|\Psi|e^{i\mathbf{qr}}$, where the vector $\mathbf q$ is aligned along the vector $[\mathbf n\times\mathbf h]$.\cite{edel1,Mineev-JETP-94}  

Choosing the $x$ axis along the intrinsic direction
$[\mathbf n\times\mathbf h]$ we find the corresponding GL equations
\begin{multline}
a\Psi-\beta|\Psi|^{2}\Psi-\xi^{2}a\mathbf D^{2}\Psi \\
+2\varepsilon_{1}h\,\mathbf D_{x}\Psi
+2\varepsilon_{3}h\,\mathbf D_{x}\mathbf D^{2}\Psi=0 .
\label{Eq_GLeq}
\end{multline}
To exclude the linear SOC term we put $\Psi=\Psi_0\,\psi\,e^{iqx}$ with 
$q=\left(a\xi^2-\sqrt{a^2\xi^4-6h^2\varepsilon_1\varepsilon_3}\right)/(3h\varepsilon_3)$ and $\Psi_0=\sqrt{a'/\beta}$, where $a'=a+2\varepsilon_{1}hq-\xi^{2}a q^{2}-2\varepsilon_{3}q^{3}$.
Finally we get
\begin{multline}
a'\psi-a'|\psi|^{2}\psi
+(\xi^{2}a+2\varepsilon_{3}hq)\mathbf D^{2}\psi
+4q\varepsilon_{3}h\,\mathbf D_{x}^{2}\psi \\
-2\varepsilon_{3}h\,\mathbf D_{x}\mathbf D^2\psi = 0.
\label{Eq_GLeq1}
\end{multline}
Restricting ourselves to the terms linear in parameters $\varepsilon_1h$ and $\varepsilon_3h$  we find approximate expressions for $q=\varepsilon_{1}h/(a\xi^{2})$, $a'=a$. 
Thus, we obtain the GL equation in the  form:
\begin{equation}
\psi-|\psi|^{2}\psi-\xi^2\mathbf D^{2}\psi
-2\gamma\,\xi^3\mathbf D_{x}{\mathbf D}^{2}\psi=0 , 
\label{Eq_GLfin}
\end{equation}
where 
$\gamma=\varepsilon_{3}h/(a\xi^{3})$.

For an isolated vortex at large distances $r\gg\xi$ from the vortex center (beyond the vortex core) we find the current distribution
\begin{align}
\mathbf j=\frac{c\Phi_{0}}{8\pi^{2}\lambda^{2}\xi}
          \left(\frac{\pmb\theta_{0}}{\tilde r}+\gamma\mathbf J\right),
\end{align}
where 
\begin{align}
\mathbf J\simeq\frac{2\ln\tilde r}{\tilde r^{2}}
               \bigl(\pmb r_{0}\cos\theta+\pmb\theta_{0}\sin\theta\bigr) \ ,
\end{align}
 and  $\tilde r=r/\xi$.

\begin{figure}[t]
\centering
\includegraphics[width=0.95\linewidth]{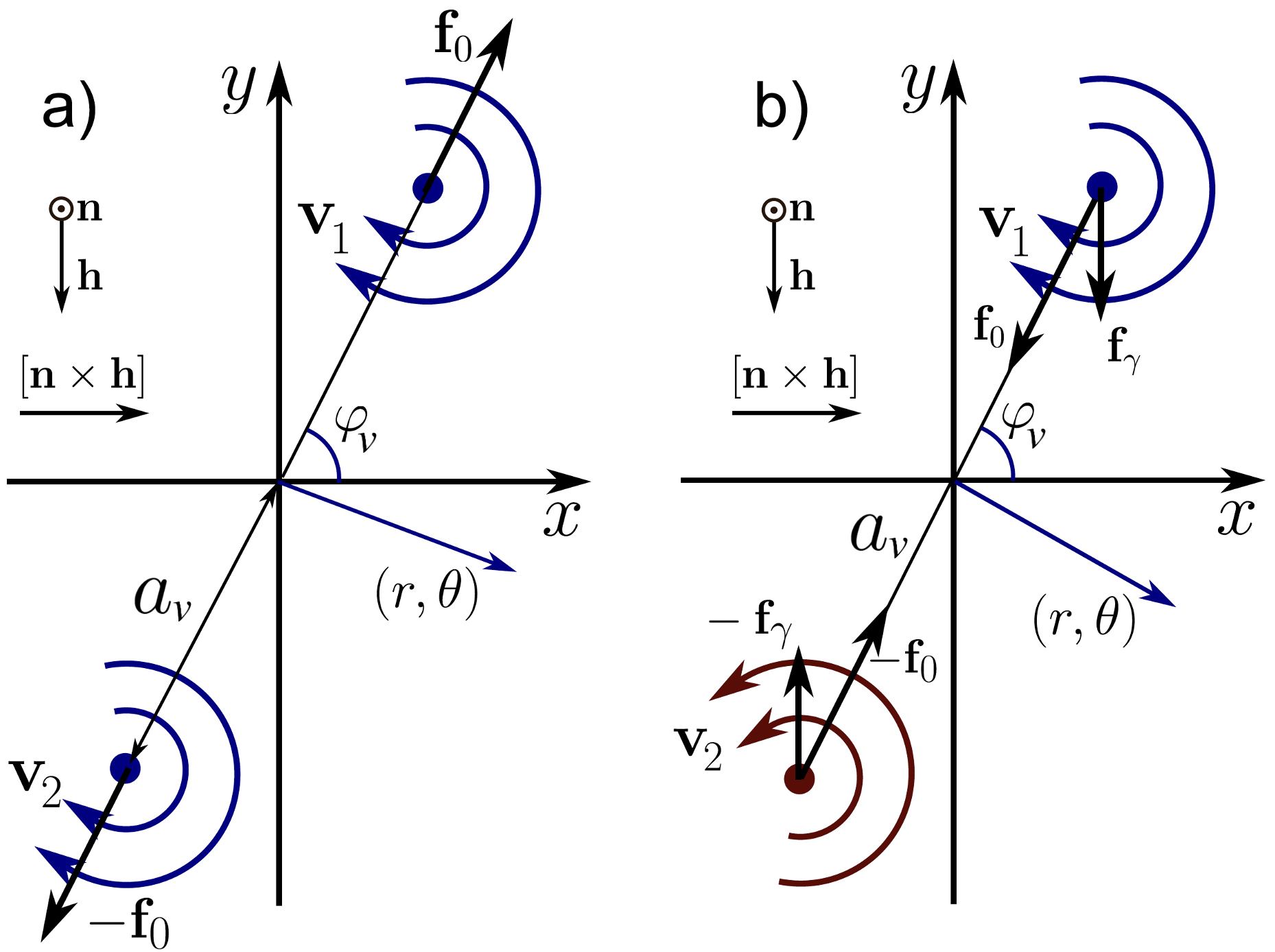}
\caption{Schematic geometry used to analyze the intervortex interaction. We introduce here a polar coordinate system ${\bf r}=(r,\theta)$, the coordinate origin is placed midway between the two vortices. a) Two vortices with coinciding vorticity which generates superfluid velocities $\mathbf v_{1}$ and $\mathbf v_2$. b) Analogous configuration of two vortices with opposite vorticities. The conventional intervortex interaction forces $\mathbf f_0$ and $\gamma$-dependent non-central forces $\mathbf f_\gamma$ are shown. Reproduced from [A. V. Putilov, D. V. Zakharov, A. Kudlis, A. S. Mel'nikov, A. I. Buzdin, Phys. Rev. B \textbf{112}, 134507 (2025)].}
\label{Fig_V_AV}
\end{figure}

As a next step we analyze the interaction energy for a vortex pair 
choosing
the vortex centers at the points $(a_v,\varphi_v)$ and $(a_v,\varphi_v+\pi)$ in polar coordinates, as it is shown in Fig.~\ref{Fig_V_AV}.  
%A velocity field $\mathbf v=\pmb\theta/r$ corresponds to the vorticity $+1$, while the reversed flow $\mathbf v=-\pmb\theta/r$ has the vorticity $-1$.  
For coinciding vorticities (Fig.~\ref{Fig_V_AV}a) the first-order $\gamma$-correction to the vortex-vortex interaction potential vanishes. In contrary, for a vortex-antivortex pair (Fig.~\ref{Fig_V_AV}b) we get
this $\gamma$-dependent correction and the related  force acting on a vortex 
assuming $\ln(a_v/\xi)\gg 1$:
\begin{equation}
\mathbf f=\mathbf f_{0}
          -4\mathbf y_0dH_c^2\xi^{3}\gamma\frac{\ln a_v}{a_v^{2}}          
          =\mathbf f_{0}+\mathbf f_\gamma,
\end{equation}
where $\mathbf f_{0}$ is the standard attractive force, and $\mathbf f_\gamma$ 
is an additional \textit{non-central} contribution to the force rotating the pair about the pair midpoint.

%%%%%%%%%%%%%%%%%%%%%%%%%%%%%%%%%%%%%%%%%%%%%%%%
%%%%%%%%%%%%%%%%%%%%%%%%%%%%%%%%%%%%%%%%%%%%%%%%

This modification of the vortex-antivortex interaction potential should also contribute to the 
 vortex interaction with the boundary which can be viewed as the interaction with an image antivortex.
In our approximation the correction to the critical current $j_c$ of the vortex entry appears to be linear in $\gamma $
%\begin{equation}
%\label{sign}
%\frac{j_c-j_c(\gamma=0)}{j_0}  \sim  \pm\gamma  
%\end{equation}
and its sign depends on the signs of the constant $\varepsilon_3$ and of the mixed product $[\mathbf n\times\mathbf h]\cdot\mathbf j$. We estimate  the value of this diode effect in critical current to be of the order of tens of percent (see [\onlinecite{Gaggioli-PRR-2024}] and references therein).

%%%%%%%%%%%%%%%%%%%%%%%%%%%%%%%%%%%%%%%%%%%%%%%%
\begin{figure}[t!]
\includegraphics[width=0.9\linewidth]{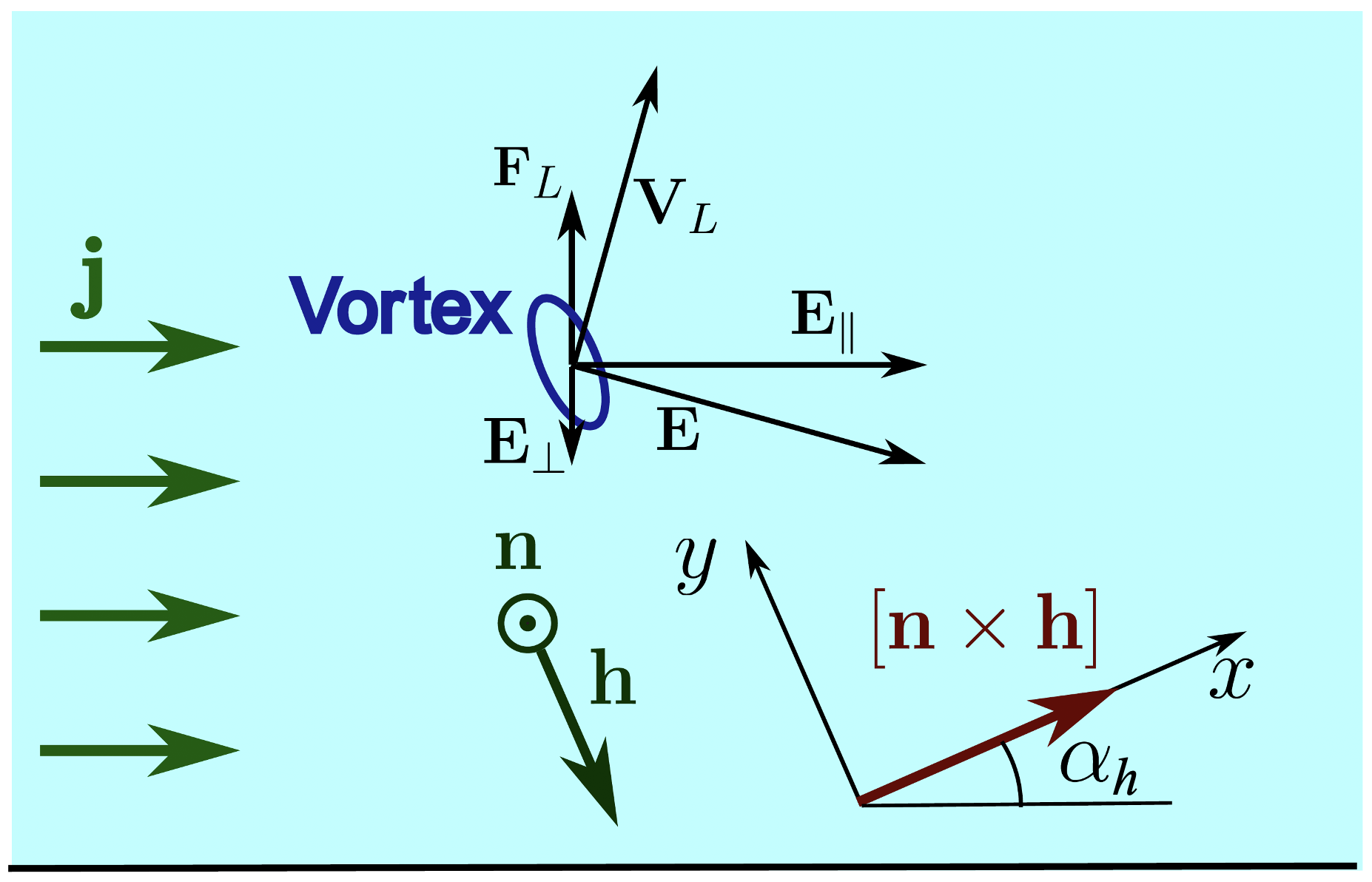}
\caption{Schematic picture illustrating the motion of a single vortex in a superconducting film in the presence of transport current  $\mathbf j$. $\mathbf F_L=c^{-1}\Phi_0[\mathbf j\times\mathbf z_0]$ is a Lorentz force, $\mathbf V_L$ is the vortex velocity and $\mathbf E$ is an induced electric field. Reproduced from [A. V. Putilov, D. V. Zakharov, A. Kudlis, A. S. Mel'nikov, A. I. Buzdin, Phys. Rev. B \textbf{112}, 134507 (2025)].}
\label{fig:hall}
\end{figure}
%%%%%%%%%%%%%%%%%%%%%%%%%%%%%%%%%%%%%%%%%%%%%%%%

From Eq.~(\ref{Eq_GLeq1}) we also see that the SOC term results in the appearance of the 
 anisotropic effective mass tensor $\hat m$
distorting the shape of vortex core.
This elliptical core distortion is determined by the product $\varepsilon_1\varepsilon_3$ and can cause a number of well known consequences for experimentally measurable quantities.\cite{Blatter-RMP-1994,Gorkov-UFN-1975,Kopnin-book-2001}
As an example, we can consider the
related effect on the vortex dynamics in the  flux-flow regime. 

For the $x$ axis chosen along the $[\mathbf n\times\mathbf h]$ direction (Fig. \ref{fig:hall}) the tensor $\hat m$ becomes diagonal with $m_x=m_0(1-3\mu_m), m_y=m_0(1-\mu_m)$, where $m_0$ is an unperturbed mass (for $\varepsilon_3=0$) and $\mu_m=2\gamma\varepsilon_1 h/(a\xi)\approx 2\gamma q\xi$.
%The mass anisotropy in given by the ratio $m_x/m_y=1-2\mu_m$, which results in the anisotropy of the vortex core profile. 
The equation of the vortex motion in a transport current ${\mathbf j}$
reads:
\begin{equation}
\hat\eta\mathbf V_L=\frac{\Phi_0}{c}[\mathbf j\times\mathbf z_0],
\end{equation}
where $\mathbf V_L$ is the vortex velocity and $\hat\eta$ is a viscosity tensor which can be found, e.g., using the results of the time dependent GL theory (see Refs.~[\onlinecite{Gorkov-UFN-1975}-\onlinecite{GenkinJETP}]).
The resulting averaged electric field is given by the expression
\begin{equation}
\mathbf E=\frac{\Phi_0B_z}{c^2}\left(
\frac{\mathbf x_0j_x}{\eta_y}+\frac{\mathbf y_0j_y}{\eta_x}
\right) \ ,
\end{equation}
where $B_z$ is the average magnetic field  along the $z$ axis. 
%In isotropic case the viscosity tensor is given by the relation $\eta_{ij}=\Phi_0^2\sigma_n/(2\pi c^2\xi^2)\delta_{ij}$, where $\sigma_n$ is a normal state conductivity and $\delta_{ij}$ is the Kronecker symbol. 
The anisotropy of viscosity can be estimated as
$\eta_x/\eta_y-1\sim\mu_m$.\cite{GenkinJETP,Hao-IEEE-1991}
For the vortices moving in a current carrying strip (see Fig. \ref{fig:hall}) this anisotropy can result in the appearance of voltage in the direction perpendicular to the current. 
The ratio of the transverse and longitudinal electric field components is given by the expression:
\begin{equation}
    E_\perp /E_\parallel\sim \mu_m \sin\alpha_h \cos\alpha_h,
\end{equation}
where $\alpha_h$ is the angle between the exchange field and current direction.
%The same value of anisotropy should reveal itself in the measurements of longitudinal and transverse voltages in  standard transport experiments.
For $q\xi\sim\gamma\sim 0.1$ we obtain a measurable value of the anisotropy parameter $\mu_m\sim 0.01$. \rev{Note that the predicted viscosity anisotropy may provide a contribution to the recently observed anisotropy of resistance in Ni/Bi bilayered superconducting system with strong SOC and exchange interaction.\cite{R2_10_1} }

\section{\label{sec:level3} Interfacial Spin-Orbit Coupling: Helical Phases and Nonreciprocal Effects in S/F Bilayers}

The appearance of helical superconducting phases has a variety of interesting consequences typical for nonreciprocal systems.
To our mind particularly perspective direction among the existing theoretical predictions relates to the field of superconducting optoelectronics: the systems with intrinsic diode effect provide a possibility to realize the
effective nonlinear interaction between the electromagnetic radiation and supercurrents.   
Below we briefly review the recent results which point out that the helical phase can generate the
photogalvanic phenomena in superconductors\cite{mironov-diode}and provide
a tool for the controlling magnetization using the transport supercurrent in
both dc and quasistatic ac regimes.\cite{Plastovets2} 

%%%%%%%%%%%%%%%%%%%%%%%%%%%%%%%%%

Using  the London-type model, i.e. neglecting the changes in the order parameter absolute value, and introducing the superconducting phase $\varphi$ and the superfluid velocity $v = (\hbar/2m)\left(\nabla\varphi + 2\pi\mathbf{A}/\Phi_0\right)$ we may write the phenomenological nonlinear expression for the current:
\begin{equation}
j_n = Q_{nm}v_{m} e^{i(\mathbf{k}\mathbf{r}-\omega t)}+ R_{nml}v_{m} v^*_{l} + \tilde R_{nml}v_{m} v_{l} e^{2i(\mathbf{k}\mathbf{r}-\omega t)},
\end{equation} 
where the indices $n$, $m$ and $l$ indicate the projections to the coordinate exes while $Q_{nm}$, $R_{nml}$ and $\tilde{R}_{nml}$ are the certain tensors. Note that here we shifted the superfluid velocity (in other words the phase gradient) to exclude an additional linear in $v$ term which appears in the expression for the system free energy so that the free energy minimum corresponds to the zero velocity $v$.

%%%%%%%%%%%%%%%%%%%%%%%
%%%%%%%%%%%%%%%%%%%%%%%
\begin{figure}[hbt!]
\begin{center}
\includegraphics[width=1\linewidth]{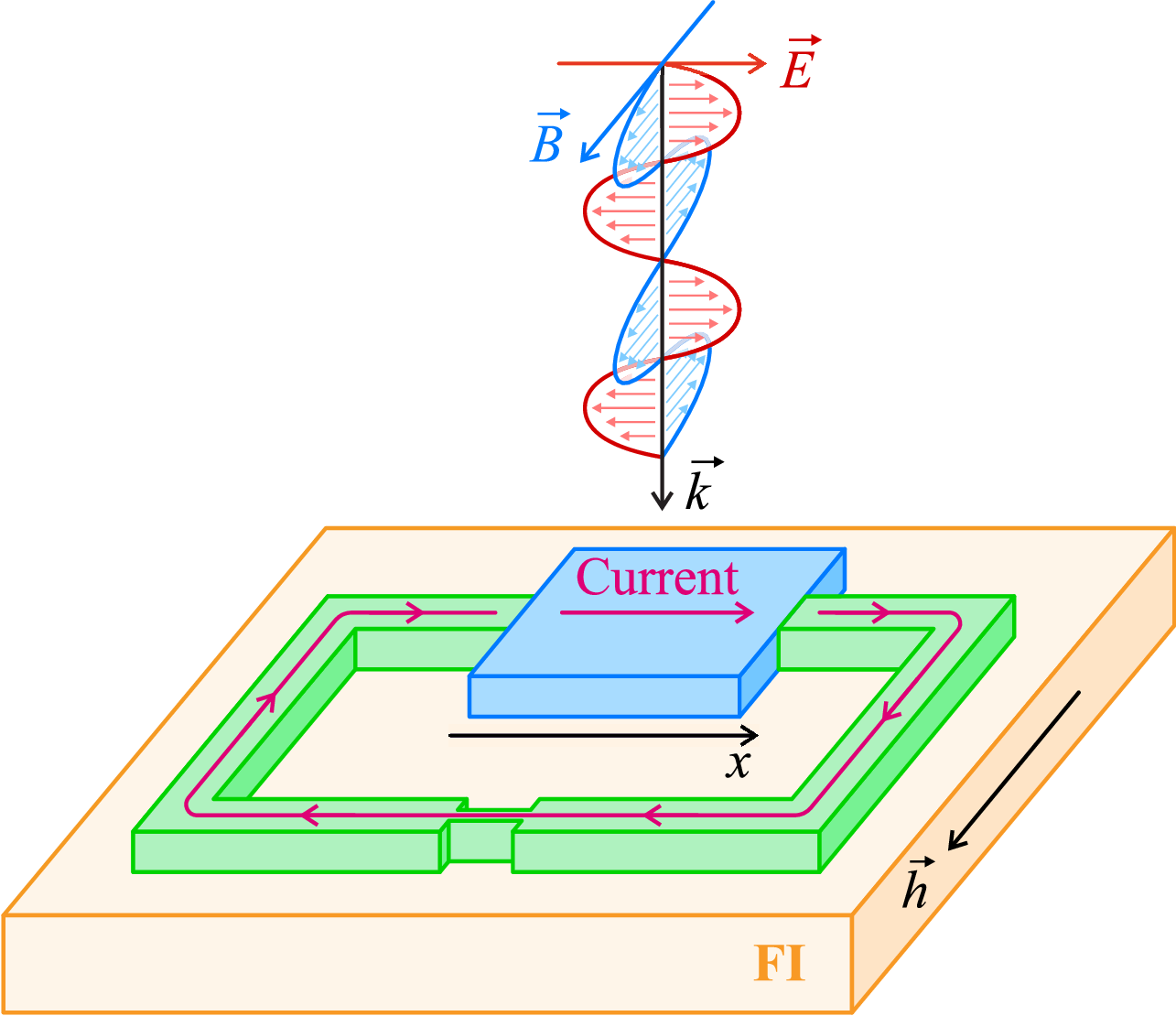}
\end{center}
\caption{Schematic setup for measurement of the rectified dc current induced by electromagnetic wave. The radiated superconducting sample with Rashba \rev{SOC} is embedded into the superconducting loop. All parts of the loop are placed on top of the ferromagnetic insulator (FI) with the in-plane exchange field. The constriction of the superconductor plays the role of weak link. Reproduced from [S. V. Mironov, A. S. Mel'nikov, A. I. Buzdin, Phys.
Rev. B \textbf{109}, L220503 (2024)].}\label{Fig_DE1}
\end{figure}
%%%%%%%%%%%%%%%%%%%%%%%
%%%%%%%%%%%%%%%%%%%%%%%

To apply this expression for the analysis of nonlinear electrodynamics and related nonreciprocal phenomena
we can consider a superconducting film of the thickness $d_s$ 
 irradiated by the linearly polarized electromagnetic wave with the wave vector perpendicular to the film surface. The thickness $d_s$ is assumed to be much larger than the interatomic distance to ensure the full electromagnetic wave reflection but, at the same time, much smaller than the London penetration depth. The latter condition allows to neglect the spatial distribution of the optically-induced electric current across the film. For  further calculations it is convenient to consider the magnetic field of the incident wave in the plane of the film in the form ${\bf B}={\rm Re}\left({\bf B}_\omega e^{-i\omega t}\right)$ where ${\bf B}_\omega$ is the complex amplitude of the wave and $\omega$ is the wave frequency. Then integrating the Maxwell equation for ${\rm curl}~{\bf B}$ over the film thickness we get:
\begin{equation}\label{Curr_omega}
2\left({\bf n}_{0}\times {\bf B_\omega}\right)=\frac{4\pi}{c}{\bf j}_{\omega}d_s,
\end{equation}
where ${\bf j}_{\omega}$ is the complex amplitude of the supercurrent at the frequency $\omega$ and the factor $2$ in the l.h.s. accounts the doubling of the amplitude of the magnetic field at the sample boundary due to the full reflection of the incident wave.
Considering the problem perturbatively we can express the superfluid velocity amplitude at the frequency of the incident wave from the relation $j_{n\omega} = Q_{nm}v_{m\omega}$ and substitute it into the expression for the rectified current
$j_{n,dc}= R_{nml}v_{m\omega} v^*_{l\omega}$.
 Further solution 
strongly depends on the proposed experimental setup and resulting boundary conditions. Indeed, if the irradiated sample is not included into the closed superconducting loop which would allow to get a circulating nonzero current  the dc photocurrent (given by the second term in the above expression) can not flow through the sample edges. The continuity of the current in this case requires the generation of the dc phase gradient which would compensate the dc photocurrent and as a result we obtain a nonzero phase difference at the edges of the sample. Thus, we get the superconducting phase battery.\cite{Robinson2019,aronov} Assuming, e.g., for simplicity the tensor $Q_{nm}$ to be diagonal we find the resulting phase difference in the form: 
$\delta\varphi\sim R_{nml}v_{m\omega} v^*_{l\omega} \sim R_{nml}j_{m\omega} j^*_{l\omega}$.

%%%%%%%%%%%%%%%%%%%%%%%
%%%%%%%%%%%%%%%%%%%%%%%
\begin{figure}[hbt!]
\begin{center}
\includegraphics[width=0.95\linewidth]{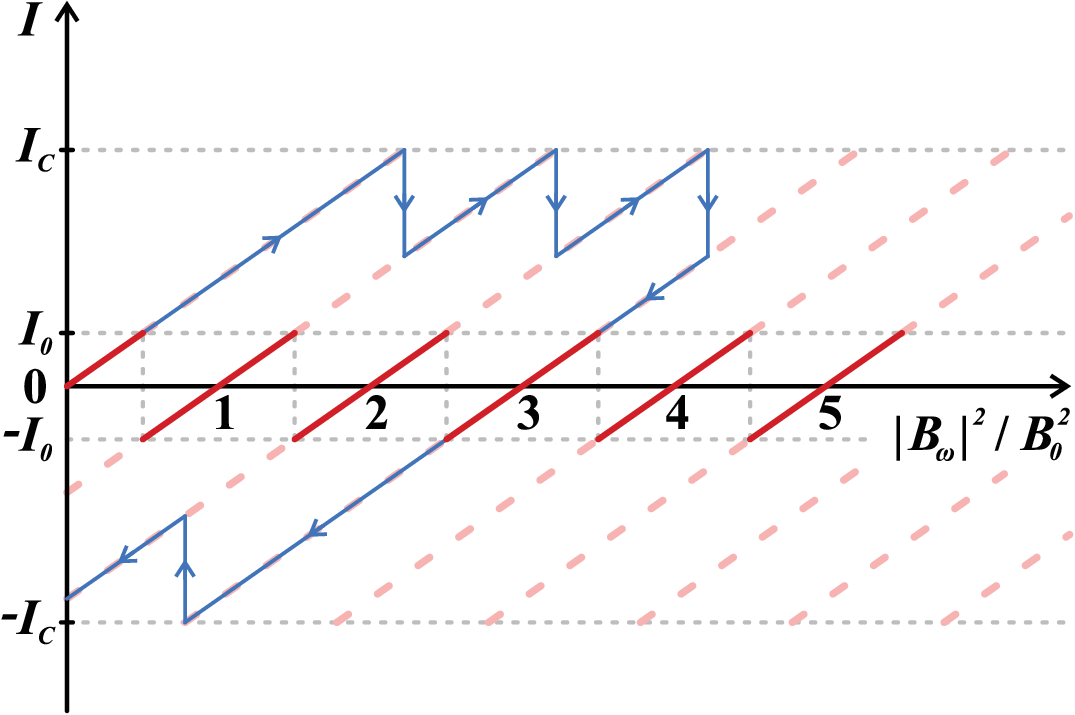}
\end{center}
\caption{Schematic plot illustrating the dependence of the rectified dc current on the intensity of the linearly polarized electromagnetic wave. Red solid lines correspond to the global energy minimum while the red dashed lines correspond to metastable states. The blue line shows an example of the hysteretic switching between the states induced by the sequential increase and decrease of the field intensity. Here $I_0=c\Phi_0/[2(L+L_k)]$, $I_c$ is the critical depairing current and $B_0$ is the certain normalization constant. Reproduced from [S. V. Mironov, A. S. Mel'nikov, A. I. Buzdin, Phys.
Rev. B \textbf{109}, L220503 (2024)].}\label{Fig_DE2}
\end{figure}
%%%%%%%%%%%%%%%%%%%%%%%
%%%%%%%%%%%%%%%%%%%%%%%

If, in opposite, the sample is included into the closed superconducting loop (this situation is sketched in Fig.~\ref{Fig_DE1}) the dc photocurrent generates the circulating supercurrent and the flux in this loop and changing the amplitude of the electromagnetic wave we can observe vortex entry/exit and switch the vorticity of the final superconducting state in the loop in a controlled manner.\cite{mironov-diode} To realize this concept it is convenient to fabricate the loop on top of the ferromagnet with the uniform in-plane exchange field and also create an additional weak link formed by a constriction or the insert of normal metal. The latter design feature ensures that the value of the maximal current which can flow through the loop without dissipation is determined by the critical current associated with this weak link. Integrating the superconducting current $I$ along the loop one finds:
\begin{equation}
I\ell =\frac{\hbar c^2S}{8\pi e\lambda^2}  
\left(\delta\varphi - 2\pi N - 2\pi \frac{\Phi}{\Phi_0}\right),
\end{equation}
where $\ell$ is the loop length, $S$ is the superconductor cross-section, $\Phi = LI/c$ is the magnetic flux through the loop, and $L$ is the geometric inductance. Introducing also the kinetic inductance of the loop $L_k=4\pi\lambda^2\ell/S$ we get
\begin{equation}\label{philoop}
I =\frac{c\Phi_0}{L+L_k}  \left(\frac{\delta\varphi}{2\pi}-N\right).
\end{equation}
Here $N$ is the integer number of vortices entering the loop one by one with the increase in the phase gain $\delta\varphi$.
This vorticity number can be determined if we consider the magnetic energy of the loop
\begin{equation}
E_N = \frac{(L+L_k)I^2}{2c^2}=\frac{\Phi_0^2}{2(L+L_k)^2} \left(\frac{\delta\varphi}{2\pi}-N\right)^2.
\end{equation}
The condition $E_N = E_{N+1}$ gives us the phase gain values $\delta\varphi = 2\pi (N+1/2)$ defining the
electromagnetic wave amplitude $B_{\omega, N+1}$ corresponding to the switching between different number of vortices in the loop. 
In Fig.~\ref{Fig_DE2} we show schematically the resulting behavior of the dc
current induced in the loop vs the electromagnetic wave intensity. Note that these transitions between different vortex states are in principle hysteretic so that experimentally 
the wave amplitudes giving the vortex entry and exit can be different (in Fig.~\ref{Fig_DE2} such hysteretic behavior is schematically shown with blue lines). 

Experimentally, to observe the predicted phenomena one may use microwave radiation which intensity would control the vorticity of the superconducting state in the loop. Taking, e.g., $c\left|B_\omega\right|^2/8\pi\sim 10~ {\rm \mu W}/{\rm \mu m}^2$ and considering the superconducting film with $d_s\sim\xi_0\sim 100~{\rm nm}$, the first critical magnetic field $B_{c1}\sim 10^{-2}~ {\rm T}$ and the maximal but reasonable values for \rev{SOC} constants\cite{mironov-diode} one may get the estimate for the superconducting phase gradient $\left|\nabla\varphi\right|\sim 0.1~ {\rm \mu m}^{-1}$. Thus, for the loops with the perimeter of several microns the radiation-controlled switching between the states with different vorticities seems to be possible.

It is interesting to note that the above phase battery effect can appear not only due to the rectification of the electromagnetic wave field but also due to the precession of the magnetic moment in the ferromagnetic layer. The latter mechanism can provide a new possibility to arrange the interaction of the supercurrents with magnonic excitations which may be of interest for coupling
the magnonic and RSFQ devices.

\section{\label{sec:ISGE} Inverse spin-galvanic phenomena in systems with spin-orbit coupling}

Finally, let us discuss the physical consequences of the electromagnetic proximity effect in superconducting systems where the electrons are subjected to the simultaneous effect of exchange field ${\bf h}$ and strong Rashba \rev{SOC} coming from the broken inversion symmetry in the direction along the certain vector ${\bf n}$. In addition to the direct  magneto-electric phenomena including the generation of spontaneous electric currents flowing parallel to the vector ${{\bf n}}\times {{\bf h}}$ discussed in Sec.~\ref{sec:level4}, these systems also support the inverse magneto-electric effects, namely, the formation of states characterized by nonzero spin polarization in the direction parallel to the vector ${\bf n}\times {\bf j}$ where ${\bf j}$ is the external current. Note that similar effects arise in semiconducting systems 
where the combination of SOC and transport currents becomes responsible for the spin Hall effect revealing itself in the generation of the spin
currents\cite{SpinHall_rev} as well as for the formation of the spin
polarized states.\cite{Ivchenko, Edelstein_1990, Aronov, Golub, Ganichev_rev, Ivchenko_book} The latter phenomenon is often referred to as inverse spin galvanic effect, Edelstein effect or current induced spin polarization (further we will use the first term).

There are several mechanisms resulting in the appearance of the inverse spin galvanic effect in superconducting systems. The first one was proposed by V. M. Edelstein who demonstrated that in the presence of the helical states induced by the interplay between SOC and magnetic field the spin structure of the wave function of Cooper pairs carrying superconducting current becomes modified in a way that the superconducting condensate acquires nonzero magnetic both in clean and dirty
limits.\cite{edel2, Edelstein_2} Another mechanism is realized in Josephson
$\varphi_0$ junctions where the transport d.c. current gives rise to the magnetic moment generation\cite{Konschelle_Bergeret} as well as to the appearance of effective magnetic field acting on the electron spins  and resulting in the precession of magnetic moment.\cite{Konschelle, Shukrinov_rev} In addition, it was shown that the formation of the long-range triplet correlations stimulated by the transport current\cite{Bobkova-PRB-2017, Silaev} also can become the source of the inverse spin-galvanic effect, which may provide the explanation for the puzzling paramagnetic response observed in experiments on the Pt/Nb systems.\cite{Lee_exp} 

Interestingly, even simpler mechanism of inverse spin-galvanic effect is related to the electromagnetic proximity effect.\cite{Mironov_ISGE} The basic system supporting this mechanism consists of a superconductor placed in electric contact with a paramagnetic (PM) material with strong SOC. The proximity effect in this system results in the formation of superconducting correlations inside the paramagnet which become involved in the generation of the Meissner currents provided the S/PM system is subjected to the external magnetic field. In the presence of SOC these currents produce the nonzero magnetization ${\bf M}$ in the paramagnetic subsystem which back-act on the Meissner currents. As a result, the magnetic field inside the S/PM system acquires the first and second-order corrections over the SOC constant manifesting inverse spin-galvanic effect. Note that the described mechanism may provide an alternative explanation of the experiment [\onlinecite{Lee_exp}] where the  Pt layer reveals the properties of the Stoner enhanced paramagnet.

%%%%%%%%%%%%%%%%%%%%%%%
%%%%%%%%%%%%%%%%%%%%%%%
\begin{figure}[hbt!]
\begin{center}
\includegraphics[width=0.8\linewidth]{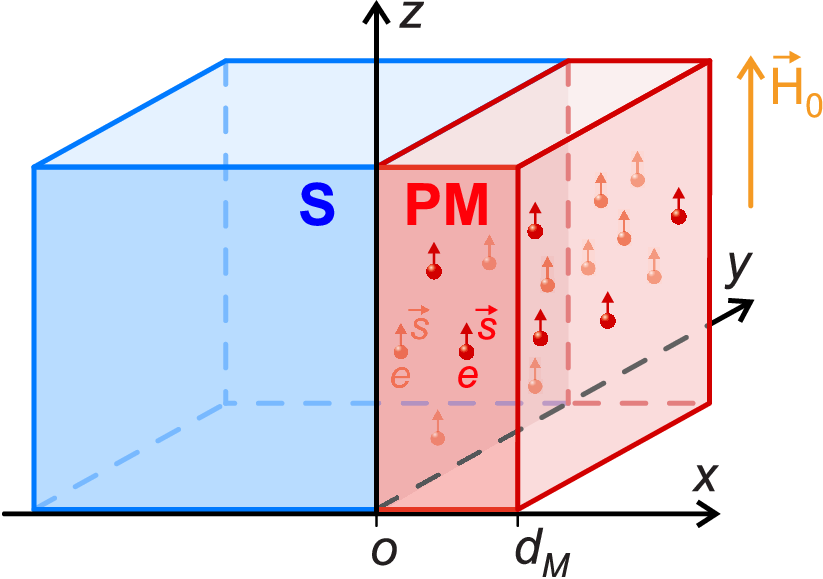}
\end{center}
\caption{Sketch of the superconductor/paramagnet heterostructure in
an external magnetic field ${\bf H}_0$ where the strong SOC gives rise to the inverse spin-galvanic effect. Reproduced from [S. V. Mironov, A. S. Mel'nikov, A. I. Buzdin, Appl. Phys. Lett. \textbf{124}, 252601 (2024)].}\label{Fig_ISGE_sketch}
\end{figure}
%%%%%%%%%%%%%%%%%%%%%%%
%%%%%%%%%%%%%%%%%%%%%%%

To provide a rigorous description of this peculiar mechanism we consider a planar heterostructure shown in Fig.~\ref{Fig_ISGE_sketch}. It consists of a thick superconductor which is put in electric contact with the PM layer of the thickness $d_M\ll\xi_n$. The electron spectrum in the paramagnet is assumed to reveal strong SOC while inside the S layer we assume the absence of SOC. The $x$ axis is chosen to be perpendicular to the S/PM interface so that the regions $x<0$ and $0<x<d_M$ correspond to the S and PM layers, respectively, while the direction of the $z$ axis is chosen along the external magnetic field $\mathbf{H}_0$. To simplify the calculations we restrict ourselves to dirty limit case which ensures that the relation between the superconducting current and the vector potential is of the London type (\ref{local}). Also we assume the temperature $T$ to be well below $T_c$ so that superconductivity is well-developed. This assumption together with the condition $d_M\ll\xi_n$ ensures that the London penetration depth $\lambda$ is almost constant across the system.

The thermodynamic potential describing the system at fixed values $T$ and $H_0$ has the form $F=S\int_{-\infty
}^{d_{M}}f(x)dx$, where 
\begin{equation}
f(x)=\frac{\left( B_{z}-H_{0}\right) ^{2}}{8\pi }+\frac{A_{y}^{2}}{8\pi
\lambda ^{2}}+\left( \alpha_{SG} M_{iz}A_{y}+\frac{\beta }{2}M_{iz}^{2}-M_{iz}B_{z}%
\right).  \label{F_def}
\end{equation}%
In Eq.~(\ref{F_def}) $S$ is the surface area of the layers, $\alpha_{SG} $ is the constant characterising the strength of SOC inside the paramagnet, $\mathbf{B}=\mathrm{curl~}\mathbf{A}$ is the magnetic field. The total magnetization ${\bf M}=\left({\bf B}-{\bf H}_0\right)/4\pi$ inside the PM layer contains two contributions: ${\bf M}={\bf M}_s+{\bf M}_i$. The first one ${\bf M}_s$ is determined by the superconducting currents while the second one ${\bf M}_i$ corresponds to the  spin polarization of itinerant electrons in the PM layer.  The constant $\beta=4\pi+1/\chi$ is related to the magnetic susceptibility $\chi$ of itinerant electrons in the normal state (i.e. in the case ${\bf M}_s=0$): ${\bf M}_i=\chi {\bf H_0}$. The stability of the paramagnetic state requires the susceptibility $\chi$ to be positive which gives the condition $\beta>4\pi$. The term proportional to the vector potential $A_{y}$ comes from the Rashba SOC inside the paramagnet. The origin of this term is the invarisnt $\sim \Psi ^{\ast }\left( {\bf n}
\times {\bf h}\right) \left( -i \nabla +2\pi{\bf A}/\Phi_0\right) \Psi $ in the Ginzburg-Landau free energy (here $\Psi$ is the superconducting order parameter).\cite{Samokhin_2004, Helical2} Note that it is this term $\propto A_{y}$ which reflects the generation of magnetization by the superconducting current (proportional to ${\bf A}$) and gives rise to the inverse spin galvanic effect. The constant $\alpha_{SG}$ is determined by the ratio between the spin-orbit velocity $u$ entering the tensor $u_{nm}$ in Eq.~(\ref{Eq3}) and the Fermi velocity $v_F$. The estimate gives
\begin{equation}\label{alpha_est}
\alpha_{SG} \sim  \frac{u}{v_F}\frac{h_0}{T_c} \frac{\Phi _{0}}{M_{0}\xi
_{0}\lambda ^{2}},
\end{equation}
where $h_{0}$ and $M_{0}$ are the exchange field and magnetic moment in the saturation regime, respectively. Interestingly, the constant $\alpha_{SG}$ is very sensitive to the system temperature since $\alpha_{SG}\propto\lambda^{-2}\propto\left|\Psi\right|^2\propto\left(T_c-T\right)$. As a consequence, the contributions to the magnetization and magnetic field coming from the SOC have peculiar temperature dependencies and, therefore, can be identified among other similar phenomena.

The first manifestation of the inverse spin galvanic effect in S/PM systems is the presence of the contribution in the magnetization $M_{iz}$ proportional to $A_y$, i.e. to the superconducting current.  Varying the functional (\ref{F_def}) with respect to $M_{iz}$ we get  
\begin{equation}
M_{iz}=\frac{1}{\beta }B_{z}-\frac{\alpha_{SG} }{\beta }A_{y},  \label{Eq_M}
\end{equation}%
where the first term stands for the usual paramagnetic response of the paramagnet  while the second term describes inverse spin galvanic effect determined by SOC. Interestingly, since the sign of the SOC constant $\alpha_{SG}$ may be either positive or negative (depending on the peculiarities of the electron spectrum of the PM material), the second contribution in Eq.~(\ref{Eq_M}) may correspond to paramagnetic or diamagnetic response. 

The second feature is the increase of the effective London penetration depth inside the paramagnet due to the SOC. To show this one may vary Eq.~(\ref{F_def}) with respect to $A_y$ and obtain the effective London equation inside the paramagnet. The corresponding London penetration depth $\lambda_M$ takes the form
\begin{equation}\label{Lambda}
\frac{1}{\lambda_M^2}= \left(\frac{1}{\lambda^2} -\frac{4\pi\alpha_{SG}^2}{\beta}\right)\frac{\beta}{\beta-4\pi}.
\end{equation}
Note that for strong SOC (specifically, for $4\pi\alpha_{SG}^2/\beta>\lambda^{-2}$) the value $\lambda_M^{-2}$ may become negative. In this case the system becomes unstable towards the spontaneous magnetic transition since for $H_0=0$ the magnetic moment is nonzero.

The solution of the London equation in the whole S/PM system in the limit $d_M\ll\lambda$ gives the spatial profile of magnetic field. Inside the S layer it has the form   
\begin{equation}
B_{z}(x)=\left( B_{0}+\Delta B\right) \exp \left( \frac{x}{\lambda }\right) ,
\label{B_res}
\end{equation}%
where $B_{0}=H_{0}\left( 1-d_{M}/\lambda \right) $ is the magnetic field at
the S/PM interface (at $x=0$) in the absence of SOC while the value 
\begin{equation}\label{dB}
\Delta B=\frac{4\pi }{(\beta -4\pi)}\left( \alpha_{SG}^2 \lambda
-\alpha_{SG}\right)d_{M} H_{0}
\end{equation}
is the correction to the magnetic field determined by the inverse spin
galvanic effect. 

The expression (\ref{dB}) for the excess magnetic field $\Delta B$ may provide some hints to the explanation of several puzzling experiments, namely, anomalous paramagnetic response observed in Pt/Nb structures\cite{Lee_exp} as well as the temperature-induced switching between diamagnetic and paramagnetic response in clean Nb cylinders  coated with Ag layer.\cite{Visani} To show this let us analyze the dependence of $\Delta B$ on the SOC constant as well as on the system temperature in more detail. The first term in the brackets of Eq.~(\ref{dB}) always provides a paramagnetic contribution while the sign of the second term depends on the sign of $\alpha_{SG}$. The resulting scenarios for the behavior of magnetic response strongly depend on the sign of $\alpha_{SG}$. For $\alpha_{SG}<0$ the value $\Delta B$ always corresponds to the
paramagnetic screening. This result may provide the explanation for the experiment [\onlinecite{Lee_exp}] with Pt/Nb structures.  \rev{Note that recently the electromagnetic proximity effect was shown to contribute to the giant demagnetization phenomenon affecting the ferromagnetic resonance (FMR) frequency in S/F/S structures.\cite{Mironov_demag} In this context, the results reviewed in this section may be relevant for the recent FMR experiments on Pt/Nb/Ni$_{80}$Fe$_{20}$/Nb/Pt structures.\cite{R1_3_1}} 

The dependencies of the functions entering the expression for the value $\Delta B$ on temperature read $\propto\left(T_c-T\right)$ and $\propto\left(T_c-T\right)^{3/2}$. These dependencies strongly differ from the standard diamagnetic contribution $\propto\left(T_c-T\right)^{1/2}$ in $B_0$. As a consequence, at temperature which are slightly below $T_c$ the dominant contribution in the response corresponds to the diamagnetism while at lower temperatures the paramagnetic contribution may strongly attenuate it. In the opposite case $\alpha_{SG}>0$ for $\alpha_{SG} \lambda
<1$ ($\alpha_{SG} \lambda>1$) the SOC gives rise to the diamagnetic (paramagnetic)
response. Since $\alpha_{SG} \propto \left( T_{c}-T\right) $ one gets $\alpha \lambda \propto \left( T_{c}-T\right) ^{1/2}$. Consequently, just below $T_{c}$ the response should be diamagnetic, while lowering the temperature may result in the crossover from the diamagnetic to the paramagnetic response. This result reminds the puzzling experimental data obtained in Ref.~[\onlinecite{Visani}] for Nb/Ag cylinders. We may speculate that the temperature induced sign change of magnetic response in these systems results from the effect of SOC.

Note that alternatively, the sign of magnetic response can be controlled by voltage at a fixed temperature. To realize this scenario one needs, e.g., fabricate the heterostructure where the PM layer is covered by a normal metal (N) separated from the paramagnet with the thin insulating (I) layer. The resulting S/PM/I/N systems provides the possibility to control the sign of the constant $\alpha_{SG}$ and, thus, the sign of Meissner response by a voltage applied between S and N layers.

Also the changes in the parameter $\alpha_{SG}\lambda$ may give rise to the controllable reversal of the magnetization in the PM layer. Thus, varying the temperature or voltage across the system is a way that the value of the parameter $\alpha_{SG}\lambda$ becomes grater than $1$ one may switch the magnetization direction and make it opposite to the one realized in the absence of the SOC.

\section{\label{sec:level5} Conclusion}

In conclusion, magnetoelectric effects in superconductor-ferromagnet
heterostructures represent a rich and promising frontier in condensed matter
physics. The interplay of superconductivity, magnetism, and \rev{SOC} provides a direct route to generating and controlling dissipationless
supercurrents and magnetic states. This synergy holds immense potential for
the development of novel superconducting spintronic devices. The ability to
control magnetic moments with non-dissipative supercurrents could pave the way
for energetically efficient memory elements, offering a new platform for
future quantum computing architectures. Continued exploration of these
phenomena promises to yield both fundamental insights and transformative
technological applications.

Considering the perspective applications in more details it is of particular importance to
suggest possible mechanisms of tuning both the magnetic ordering and the Rashba \rev{SOC}. \rev{The gate-controlled, time-dependent spin-orbit coupling in a Josephson junction opens a way to strongly modify current-phase relations and provides a mechanism for dynamically driving the junction, even in the absence of a bias current.\cite{R2_13_1}} As a very promising route, offering a
non-dissipative alternative to conventional spintronic control mechanisms we can mention, e.g.,
the mechanism of the active manipulation and
switching of magnetic moments associated with the interaction between the current generated by the magnetic dot and
transport supercurrent studied in Ref.~[\onlinecite{Chudnovsky}].
The superconducting current can be also used for manipulation of the skyrmion structures through the above
mechanisms of coupling between magnetic skyrmions and
 vortices  which highlights the potential for dissipationless manipulation in these systems.

The convergence of Rashba \rev{SOC} and superconductivity offers a sophisticated pathway for 
electrostatic control of the quantum states. Since the anisotropic Rashba SOC is intrinsically linked to the structural 
inversion asymmetry, it can be modulated by an external electric field. This tunability allows for the direct coupling 
of superconducting phenomena, such as the superconducting diode effect and spontaneous current generation, to the gate 
voltages.\cite{Bordoloi,Liang} 
By leveraging this gate-tunable SOC, it becomes possible to design novel functionalities where magnetic switching and 
phase-coherent transport are managed via the low-power electrostatic control.

The consideration of phenomena caused by the interface SOC can be naturally generalized for the case of momentum dependent exchange fields
which can appear even in the absence of relativistic SOC effects. This issue is particularly interesting in the context of recent studies
of so-called altermagnetic materials. \rev{It is interesting to note that the first hints of momentum-dependent spin-splitting in the antiferromagnetic MnTe (now a well-established altermagnet) were obtained using the Green's function method more than $40$ years ago.\cite{Sandratskii}} The discovery of this new class of magnetic materials suggests interesting perspectives for development of spintronics.\cite{bai,jung} 
Altermagnets are known to exhibit zero-net
magnetization and simultaneously rather strong Zeeman-like spin splitting of the
electron spectrum\cite{smej} with the sign alternating over the Brillouin zone.
Proximity effect in superconductor - altermagnet structures can be less efficient
in suppressing the superconducting ordering than in SF systems,\cite{chour,vas1,vas2}
which allows to consider altermagnets as possible material base for superconducting
spintronics. \rev{Recently, first-principles calculations of the proximity effect between a superconductor and an altermagnet were reported in [\onlinecite{R2_14_2}} The use of these materials provides, e.g., interesting new variants of 
superconducting diode effect,\cite{sahoo,deb,chak} modification of vortex matter physics\cite{mazanik} as well as engineering of Cooper pair wave function including both its orbital and spin structure.

The studies of the interaction of electromagnetic waves with S/F systems open another perspective and almost unexplored direction of superconducting spintronics: physics of light control of triplet superconductivity and light-operated Josephson transport.
Superconductor/ferromagnet/superconductor Josephson junctions featuring non-collinear magnetic textures in the 
barrier layer facilitate the generation of spin-polarized triplet supercurrents. This phenomenon serves as a 
cornerstone for superconducting spintronics and may provide a framework for light-guided S/F/S junctions, focusing on the 
non-equilibrium dynamics of the superconducting order parameter when subjected to ultra-short laser pulses.
The superconducting response to femtosecond-scale reorientations of the exchange field will address the fundamental 
challenge of ultrafast perturbations in quantum condensates. The localized manipulation of magnetization within the 
ferromagnetic weak link via a focused optical beam provides a surgical tool for generating odd-frequency triplet 
correlations which provides the basis of development of the essential building block for dissipationless spintronic logic.
Finally, we note that additional promising research trajectories in S/F optoelectronics can include 
the ultrafast modulation of S/F proximity effects using the circularly polarized light and the related inverse Faraday effect responsible for switching both the magnetic texture and the supercurrent states.\cite{Kimel_1, Kimel_2, Kimel_3, Kimel_4, Sheldon, Mironov_IFE, Croitoru_1, Croitoru_2, Plastovets_IFE, Croitoru_3, Putilov_IFE, Dzero, Balatsky, Zuev, Mironov_IFE_rev} 

All these examples and developments provide convincing evidence that the recent progress in physics of S/F heterostructure not only challenges our understanding of the interplay between supercondutivity and magnetism but also
opens new avenues for designing novel, energy-efficient electronic devices.

\begin{acknowledgments}

The authors thank A. A. Kopasov for technical help with manuscript preparation. This work was supported by the Russian Science Foundation (Grant No. 25-
12-00042) in part of the analysis of photogalvanic and vortex physics phenomena (Sec.~\ref{sec:level4} and Sec.~\ref{sec:level3}), and by the Grant of the Ministry of
science and higher education of the Russian Federation
No. 075-15-2025-010 in part of the study of electromagnetic proximity effect (Sec.~\ref{sec:EPE}). A.I.B. acknowledges support by LIGHT S\&T Graduate Program, GPR LIGHT and ANR SUPERFAST in part of the study of the inverse spin-galvanic effect (Sec.~\ref{sec:ISGE}). S.V.M. acknowledges the financial
support of the Foundation for the Advancement of Theoretical Physics and Mathematics BASIS (Grant No. 23-1-2-32-1).

\end{acknowledgments}

\section*{Data Availability Statement}
%
%The data that support the findings of this study are available from the corresponding author upon reasonable request.
%
%OR

Data sharing is not applicable to this article as no new data were created or analyzed in this study.

\appendix

%\section{Appendixes}

\nocite{*}
%\bibliography{aipsamp}
% TODO: Convert bibliography into bibtex format

\end{document}